\documentclass[aps,prd,onecolumn,eqsecnum,amsmath,nofootinbib,preprintnumbers]{revtex4}%
\usepackage[dvipsnames]{xcolor}
\usepackage{color,graphicx,float,subfigure,xcolor}
\usepackage{amsfonts,amssymb,theorem,mathrsfs,times}
\usepackage{bm}
\usepackage{multirow}
\usepackage{mathtools}
\usepackage{amsfonts,amssymb,theorem,mathrsfs}
\usepackage{dsfont}
\usepackage{setspace}
\usepackage{amsmath} 
\usepackage{graphicx}
\usepackage[colorlinks,linkcolor=blue,anchorcolor=blue,citecolor=blue]{hyperref}   
\usepackage{ulem}   
\usepackage[english]{babel}
\usepackage{float}
\usepackage{appendix}
\usepackage{cleveref}
\usepackage{verbatim}

{\theorembodyfont{\upshape}
	}
{\theorembodyfont{\upshape}
	}
{\theorembodyfont{\upshape}
	}
{\theorembodyfont{\upshape}
	}
{\theorembodyfont{\upshape}
	}
{\theorembodyfont{\upshape}
	}
\addtocounter{MaxMatrixCols}{10}
\newcommand{\dalm}{\kern1pt\vbox{\hrule height 0.9pt\hbox{\vrule width
			0.9pt\hskip 2.5pt\vbox{\vskip 5.5pt}\hskip 3pt\vrule width
			0.3pt}\hrule height 0.3pt}\kern1pt}

\begin{document}
\thispagestyle{empty}
\preprint{\hfill {\small {ICTS-USTC/PCFT-25-33}}}
\title{Pseudospectrum and time-domain analysis of the EFT corrected black holes}
	
%

\author{Li-Ming Cao$^{a\, ,b}$\footnote{e-mail address: caolm@ustc.edu.cn}} 

\author{Ming-Fei Ji$^{a}$\footnote{e-mail address: jimingfei@mail.ustc.edu.cn (corresponding author)}}

\author{Liang-Bi Wu$^{c\, ,d}$\footnote{e-mail address: liangbi@mail.ustc.edu.cn} }
 
\author{Yu-Sen Zhou$^a$\footnote{e-mail address: zhou\_ys@mail.ustc.edu.cn}}		
	
\affiliation{${}^a$Interdisciplinary Center for Theoretical Study and Department of Modern Physics, University of Science and Technology of China, Hefei, Anhui 230026, China}

\affiliation{${}^b$Peng Huanwu Center for Fundamental Theory, Hefei, Anhui 230026, China}

\affiliation{${}^c$School of Fundamental Physics and Mathematical Sciences, Hangzhou Institute for Advanced Study, UCAS, Hangzhou 310024, China}

\affiliation{${}^d$University of Chinese Academy of Sciences, Beijing 100049, China}

\date{\today}
	
\begin{abstract}
We study the linear perturbations of a spherically symmetric black hole corrected by dimension-6 terms in the effective field theory (EFT) of gravity. The solution is asymptotically flat and characterized by two parameters---a mass parameter $M$ and a dimensionless parameter $\varepsilon$ related to the EFT length scale $l$, and the perturbation equation incorporates a velocity factor which is not constant. The quasinormal modes (QNMs) and time-domain waveforms are studied within the hyperboloidal framework. This approach reproduces the breakdown of the isospectrality and reveals that higher overtones are more sensitive to $\varepsilon$. As for the time domain, the mismatch function is introduced and found to scale as $\varepsilon^2$, which demonstrates that the waveform is stable as $\varepsilon$ varies. Finally, a velocity-dependent energy norm is employed to compute the pseudospectrum and characterize the migration of the QNM spectrum.  We further define a quantity $\epsilon_c$ that describes the magnitude of the instability of a QNM spectrum. Our analysis reveals that the dependence of $\epsilon_c$ on $\varepsilon$ is complicated---it may increase, decrease or even be nonmonotonic.
\end{abstract}

\maketitle
	

\section{Introduction}
Higher order derivative terms arise naturally when quantum effects are involved. In the framework of the low–energy effective field theory (EFT) of gravity~\cite{Donoghue:1994dn,Donoghue:1995cz,Burgess:2003jk,Donoghue:2012zc}, the higher-order curvature terms added to the Einstein-Hilbert action come from the unknown UV complete theory. EFT offers a systematic approach for incorporating all possible modifications to an established theory that may result from new physics. Since the EFT is valid only for the length scale above a cut off, all higher-dimensional curvature terms should be treated perturbatively.

One way~\cite{Arun:2006hn,Yunes:2009ke,Mishra:2010tp,Agathos:2013upa} to test gravity theory is through the gravitational waves (GWs), which can be created by perturbations of spacetime. The response of a black hole under perturbation can be divided into three stages~\cite{Leaver:1986gd}: an early prompt response, followed by a damped oscillatory ringdown and finally the late-time power-law decay tail~\cite{Price:1971fb,Price:1972pw}. The ringdown phase can be described as a superposition of the quasinormal modes (QNMs)~\cite{Nollert:1999ji, Kokkotas:1999bd,Berti:2009kk,Konoplya:2011qq,Berti:2025hly}, which are defined as the solutions of perturbation equations with outgoing boundary conditions.

The EFT of gravity gives an effective formalism for encoding different extensions to general relativity (GR), and provides a framework to test the quantum corrections by GWs~\cite{Endlich:2017tqa,Sennett:2019bpc}. The additional curvature terms in the EFT of gravity make a difference in the response of a black hole under perturbation. We only consider the ``Schwarzschild branch'' solutions~\cite{deRham:2020ejn,Silva:2024ffz}, in which the higher-dimensional terms give a small linear-order correction of the original Schwarzschild spacetime. For the Schwarzschild branch solutions, the dimension-$4$ curvature terms do not contribute to the background nor the perturbation equation~\cite{deRham:2020ejn}, and the leading-order correction is caused by dimension-6 terms. The deviation of this spacetime's metric from the Schwarzschild black hole's metric depends on a dimensionless parameter $\varepsilon$, which relies on the dimension-$6$ coupling constant, the mass parameter and the EFT length scale. This spacetime is an asymptotically flat black hole solution. The linear perturbation equation also differs from its GR counterpart, not only in the effective potential but also in an extra velocity factor, which is not the speed of light, except at the event horizon and the infinity. Using the corrected perturbation equation, one can calculate the corrected QNMs of the EFT corrected black holes~\cite{Silva:2024ffz}. There are many effects on the QNM spectra caused by the additional curvature terms, such as the breakdown of isospectrality ~\cite{Silva:2024ffz,Cardoso:2018ptl,deRham:2020ejn,Cano:2021myl} and an increased sensitivity of higher overtones to new length scales~\cite{Silva:2024ffz}.

With the help of the hyperboloidal framework~\cite{Zenginoglu:2007jw,Ansorg:2016ztf,PanossoMacedo:2018hab,Zenginoglu:2011jz, PanossoMacedo:2023qzp, Zenginoglu:2024bzs, PanossoMacedo:2024nkw,Jaramillo:2020tuu,Zenginoglu:2025sft,PanossoMacedo:2019npm}, the perturbation equation can be recast into first-order linear differential equations, i.e., $\partial_tu=Lu$, and the outgoing boundary condition is equivalent to the regular condition of the solution. Therefore, solving for the QNM spectra becomes equivalent to finding the eigenvalue of a non-self-adjoint linear operator $L$. The time-domain waveform can also be obtained by solving these equations. It is well established that the QNM spectrum of black holes exhibits instability~\cite{Jaramillo:2020tuu,konoplya:2022pbc,Shen:2025yiy,Xie:2025jbr,Qian:2020cnz,Daghigh:2020jyk,Liu:2021aqh,Li:2024npg,Qian:2024iaq,Berti:2022xfj,Cheung:2021bol,Yang:2024vor,Courty:2023rxk,Cardoso:2024mrw,Ianniccari:2024ysv,MalatoCorrea:2025iuc}, which was first studied by Nollert and Price~\cite{Nollert:1996rf,Nollert:1998ys}. One of the methods for studying the spectrum instability involves modifying the effective potential~\cite{Qian:2020cnz,Daghigh:2020jyk,Liu:2021aqh,Li:2024npg,Qian:2024iaq,Xie:2025jbr,Berti:2022xfj,Cheung:2021bol,Yang:2024vor,Courty:2023rxk,Cardoso:2024mrw,Ianniccari:2024ysv,MalatoCorrea:2025iuc}. When the small extra bump becomes negative enough, spacetime will even become unstable~\cite{Mai:2025cva,Lan:2025brn}, where a dark matter distribution is introduced in~\cite{Lan:2025brn} to get the negative modification of potential. In addition, one can also change the boundary conditions of QNMs to study spectra instability~\cite{Solidoro:2024yxi,Oshita:2025ibu}. Since the hyperboloidal framework recasts the QNM problem as a pure eigenvalue problem, it becomes amenable to powerful mathematical methods. Pseudospectrum analysis, as a sophisticated method, is utilized to investigate the instability of the QNM spectrum within hyperboloidal framework~\cite{trefethen2020spectra,Boyanov:2024fgc, Jaramillo:2020tuu,Destounis:2023ruj,Jaramillo:2021tmt}. It was originally developed to characterize non-self-adjoint operators and quantify how eigenvalues respond to perturbations of a given norm $\epsilon$~\cite{trefethen2020spectra}. The $\epsilon$-pseudospectrum provides visual insight into spectrum stability through its contour structure in the complex plane---open contours indicate spectral instability under perturbations. This approach has been successfully applied across various spacetimes~\cite{Jaramillo:2020tuu, Destounis:2021lum,Cao:2024oud,Sarkar:2023rhp, Destounis:2023nmb,Luo:2024dxl,Warnick:2024usx,Arean:2024afl, Cownden:2023dam, Boyanov:2023qqf,Garcia-Farina:2024pdd,Arean:2023ejh,Boyanov:2022ark,Cao:2024sot,Carballo:2025ajx,Chen:2024mon,Siqueira:2025lww,Besson:2024adi,dePaula:2025fqt,Cai:2025irl,Carballo:2024kbk}.

Based on the model in~\cite{Silva:2024ffz}, we aim to further investigate the QNMs, time-domain waveform and the pseudospectrum of the EFT corrected black hole within the hyperboloidal framework. We want to answer these questions: (a) What are the effects of incorporating dimension-$6$ terms into the action on the QNM spectrum? (b) How do the time-domain waveform depend on $\varepsilon$? (c) How does the dimension-$6$ terms affect the spectrum instability? For an EFT corrected black hole with fixed mass, the tortoise coordinate and then the hyperboloidal coordinates depends on $\varepsilon$. The dependence of the hyperboloidal coordinates and the phase velocity on $\varepsilon$ both give contribution to the linear operator $L$, so the QNM spectrum as the eigenvalue of $L/i$ depend on $\varepsilon$. In this way, we can study the effect on the QNM spectrum caused by the higher-dimensional curvature terms (which is characterized by the value of $\varepsilon$). The QNM spectrum for polar and axial parity perturbations bifurcates as $\varepsilon$ increasing from $0$; we therefore confirm the violation of isospectrality using a new method. In order to calculate the pseudospectrum for spectrum instability analysis, we need to define a norm of the linear operator $L$. We choose the modified energy norm~\cite{Jaramillo:2020tuu, Gasperin:2021kfv, Besson:2024adi} incorporating the effect of sound velocity. To our knowledge, this is the first time that the correction of sound velocity has been introduced into the energy norm. The $\epsilon$-pseudospectrum for different $\varepsilon$ illustrates the (in)stability of QNM spectrum in different EFT length scales.

This work is organized as follows. In Sec. \ref{theory_solution_perturbation}, we review the EFT of gravity together with the spherically symmetric black hole solution and the perturbation equation. Section \ref{Hyperboloidal framework} presents the hyperboloidal coordinates adopted for the EFT corrected black hole, followed by the reformulation of the perturbation equation and a test of the causal structure. In Sec. \ref{QNMs}, we calculate the QNMs based on hyperboloidal framework and study the migration of QNM spectra as the parameter $\varepsilon$ changes. In Sec. \ref{TD_wave_form},  the time-domain waveforms are calculated and their dependence on $\varepsilon$ is analyzed. Section \ref{stability_analysis} begins with a review of the definition of pseudospectrum, after which an energy norm is defined and employed to compute the pseudospectrum for the EFT corrected black hole. In Sec. \ref{conclusions}, we summarize and discuss our results. Explicit expressions for the tortoise coordinate and its asymptotic behavior are included in Appendix \ref{Appendix A}, while Appendix \ref{Appendix B} details the modifications to the effective potential induced by the EFT corrections.

\section{Dimension-6 EFT corrected black hole and its perturbations}\label{theory_solution_perturbation}
In this section, following~\cite{Silva:2024ffz}, we give a brief introduction on the low-energy effective field theory and its perturbation equations. The general structure of the action for the low-energy effective field theory of the gravity theory can be written as
\begin{eqnarray}\label{action}
    S=\frac{1}{16\pi}\int{\mathrm{d}^4x}\sqrt{-g}R+\frac{1}{16\pi}\sum_{n\geqslant 2}{l^{2n-2}S^{(2n)}}\, ,
\end{eqnarray}
where $l$ is a length scale much less than the length scale related to the black hole mass $M$, and the action $S^{(2n)}$ corresponds to the $n$th-order curvature term involving $2n$-order derivatives of the metric. In this work, we only consider vacuum solutions with $R_{\mu\nu}=0$. As shown in~\cite{deRham:2020ejn}, the dimension-$4$ operators cannot lead to any leading-order correction in perturbation equations for these solutions. Thus, the leading-order correction is contributed by the dimension-$6$ operators. Given this, the relevant action is
\begin{eqnarray}\label{6daction}
    S=\frac{1}{16\pi}\int\mathrm{d}^4x
    \Big(R+l^4\mathscr{L}\Big)\, ,
\end{eqnarray}
where $\mathscr{L}$ has a form
\begin{eqnarray}
    \mathscr{L}=g_e {R_{\mu\nu}}^{\rho\sigma}{R_{\rho\sigma}}^{\delta\gamma}{R_{\delta\gamma}}^{\mu\nu}+g_o {R_{\mu\nu}}^{\rho\sigma}{R_{\rho\sigma}}^{\delta\gamma}{\tilde{R}}_{\delta \gamma}^{\quad\mu \nu}\, .
\end{eqnarray}
Here, $\tilde{R}_{\mu\nu\rho\sigma}=(1/2){\epsilon_{\mu\nu}}^{\alpha\beta}R_{\alpha\beta\rho\sigma}$ is the dual Riemann tensor, in which $\epsilon_{\mu\nu\rho\sigma}$ is the Levi-Civita tensor, and $g_{e,o}$ are dimensionless constants. The field equation obtained from the action \eqref{6daction} is 
\begin{eqnarray}\label{EOM}
    \mathscr{E}_{\alpha\beta}=G_{\alpha\beta}+l^4\mathscr{S}_{\alpha\beta}=0\, ,
\end{eqnarray}
where
\begin{eqnarray}
    \mathscr{S}_{\alpha\beta}&=&{P_{(\alpha}}^{\rho\sigma\delta}R_{_{\beta)\rho\sigma\delta}}-\frac{1}{2}G_{\alpha\beta}\mathscr{L}+2\nabla^{\sigma}\nabla^{\rho}P_{(\alpha|\sigma|\beta)\rho}\, ,\\
    P_{\alpha\beta\mu\nu}&=&3g_e{R_{\alpha\beta}}^{\rho\sigma}R_{\rho\sigma\mu\nu}+\frac{3}{2}g_o({R_{\alpha\beta}}^{\rho\sigma}\tilde{R}_{\rho\sigma\mu\nu}+{R_{\alpha\beta}}^{\rho\sigma}\tilde{R}_{\mu\nu\rho\sigma})\, .
\end{eqnarray}

In this work, we assume that $g_o=0$ and $g_e=g$ is positive. For such conditions, the spherically symmetric black hole solution can be found in~\cite{deRham:2020ejn,Cano:2021myl,Cano:2019ore,Silva:2024ffz}, whose line element is written as 
\begin{eqnarray}\label{metric}
    \mathrm{d}s^2=-A^2B\mathrm{d}t^2+B^{-1}\mathrm{d}r^2+r^2\mathrm{d}\theta^2+r^2\sin^2\theta \mathrm{d}\varphi^2\, ,
\end{eqnarray}
where the metric functions $A$ and $B$ are
\begin{eqnarray}
    A&=&1-108\varepsilon\frac{M^6}{r^6}+\mathcal{O}(\varepsilon^2)\, ,\label{Adef}\\
    B&=&1-\frac{2M}{r}+216\varepsilon\Big(1-\frac{49}{27}\frac{M}{r}\Big)\frac{M^6}{r^6}+\mathcal{O}(\varepsilon^2)\, ,\label{Bdef}
\end{eqnarray}
and the dimensionless parameter $\varepsilon$ is defined as
\begin{eqnarray}\label{dimensionless_parameter_varepsilon}
    \varepsilon\equiv\frac{g l^4}{M^4}\, .
\end{eqnarray}
The solution reduces to the Schwarzschild black hole solution when $\varepsilon=0$, and the radius of the event horizon is given by
\begin{eqnarray}\label{rh}
    r_\text{h}=2M\Big(1-\frac{5\varepsilon}{16}\Big)+\mathcal{O}(\varepsilon^2)\, .
\end{eqnarray}

It is necessary to put constraint on the parameter $\varepsilon$ through GW events. However, since astrophysical black holes possess spin, a rotating solution to Eq. \eqref{EOM} is required, rather than the spherically symmetric case considered in this work. In~\cite{Silva:2022srr}, the authors investigated such a spinning black hole solution within an effective field theory framework. By applying a Bayesian analysis to the ringdown signals of GW150914 and GW200129, they derived an upper bound on the length scale $l$ associated with dimension-6 operators in the effective action, namely $l \leqslant 38.2\,\text{km}$.

Given that the Schwarzschild radius of the Sun is approximately $2.95\,\text{km}$, the mass $M$ can be expressed as $M \sim 1.48\, \text{km} \times M/M_{\odot}$. In the configuration of~\cite{Silva:2022srr}, where $g = 1$, the upper bound of the dimensionless parameter $\varepsilon$ defined in Eq. \eqref{dimensionless_parameter_varepsilon} satisfies  
\begin{eqnarray}
    \varepsilon_{\text{lim}}=\Big(\frac{38.2\,\text{km}}{1.48\,\text{km}\times M/M_{\odot}}\Big)^4\, .
\end{eqnarray}
 
Note that the final mass of observed binary black hole merger GW events is about $20 M_{\odot}\sim200 M_{\odot}$. For a black hole whose mass is about $20 M_{\odot}$ the upper bound $\varepsilon_{\text{lim}}$ is about $3$; for a black hole whose mass is about $200 M_{\odot}$, $\varepsilon_{\text{lim}}$ is about $3\times10^{-4}$.

Back to the spherically symmetric solution in Eq. \eqref{metric}, within the framework of this EFT theory, the perturbation equations for the metric (\ref{metric}) have been worked out in~\cite{Silva:2024ffz}. The time-domain master equations for the polar $(+)$ sector and the axial $(-)$ sector are
\begin{eqnarray}\label{tdeq}
    \Big[-\frac{1}{c_s^2(r)}\frac{\partial^2}{\partial t^2}+\frac{\partial^2}{\partial r_{\star}^2}-V_{\ell}^{(\pm)}(r)\Big]\phi_{\ell}^{\pm}(t,r)=0\, ,
\end{eqnarray}
where the tortoise coordinate $r_\star$ satisfies 
\begin{eqnarray}\label{tortoise_coordinate}
    \frac{\mathrm{d}r_{\star}}{\mathrm{d}r}=\frac{1}{AB}\, ,
\end{eqnarray}
and the domain we focus on, $r_\text{h}\leqslant r<\infty$, is mapped to $-\infty<r_{\star}<\infty$ provided that $\varepsilon\lesssim0.59$. Beyond this range, the tortoise coordinate is ill defined. The explicit expression of the tortoise coordinate is shown in Appendix \ref{Appendix A}. In Eq. (\ref{tdeq}), $V_l^{\pm}$ and $c_s$ are the effective potential and propagation velocity of the perturbations, respectively. The effective potentials of the above two sectors have the forms
\begin{eqnarray}\label{effective_potentials}
    V_l^{(\pm)}=\Big(1-\frac{r_\text{h}}{r}\Big)\left[\Bar{V}_{\ell}^{(\pm)}+\varepsilon\delta V_{\ell}^{(\pm)}+\mathcal{O}(\varepsilon^2)\right]\, ,
\end{eqnarray}
where $\Bar{V}_{\ell}^{(\pm)}$ are the well-known Zerilli and Regge-Wheeler potentials contributed by GR, which are given by
\begin{eqnarray}\label{Zerilli_Regge_Wheeler_potentials}
    \Bar{V}_\ell^{(+)}&=&\frac{1}{(r\Lambda_{\ell})^2}\Big[2\lambda_{\ell}^2(\Lambda_{\ell}+1)+\frac{18M^2}{r^2}\Big(\lambda_{\ell}+\frac{M}{r}\Big)\Big]\, ,\\
    \Bar{V}_{\ell}^{(-)}&=&\frac{1}{r^2}\Big[{\ell}(\ell+1)-\frac{6M}{r}\Big]\, ,
\end{eqnarray}
in which $\lambda_{\ell}=(\ell+2)(\ell-1)/2$, $\Lambda_{\ell}=\lambda_{\ell}+3M/r$. Other parts of $V_l^{(\pm)}$,  $\delta V^{(\pm)}$, are the modifications from EFT, whose explicit expressions can be found in Appendix \ref{Appendix B}. Finally, the square of the velocity depends on the position in the spacetime, i.e.,
\begin{eqnarray}\label{sound_velocity}
    c_s^2=1-288\varepsilon\Big(1-\frac{r_\text{h}}{r}\Big)\frac{M^5}{r^5}+\mathcal{O}(\varepsilon^2)\, ,
\end{eqnarray}
and the squared velocity $c_s^2$ tends to unity as $r\rightarrow r_\text{h}$ or $r\rightarrow \infty$, which ensures that the perturbations still propagate at the speed of light at the boundary of the computational domain, therefore, the original boundary conditions of the QNM problem will not be affected by the extra factor of the squared velocity.

\section{Hyperboloidal framework of the EFT corrected black hole}\label{Hyperboloidal framework}
The calculation of QNMs requires careful treatment of boundary conditions and the divergence of eigenfunctions at the boundaries. The hyperboloidal framework~\cite{Zenginoglu:2007jw,Ansorg:2016ztf,PanossoMacedo:2018hab,Zenginoglu:2011jz, PanossoMacedo:2023qzp, Zenginoglu:2024bzs, PanossoMacedo:2024nkw,Jaramillo:2020tuu,Zenginoglu:2025sft,PanossoMacedo:2019npm} provides a solution to these challenges. This approach employs constant-$\tau$ slices that simultaneously extend to null infinity and penetrate the event horizon, thereby naturally regularizing the eigenfunctions. Moreover, the framework's geometric structure ensures that light cones at the boundaries are oriented outward from the computational domain, automatically satisfying the required boundary conditions.

For the EFT corrected black hole (\ref{metric}), we study the compact hyperboloidal coordinates $(\tau, \sigma, \theta, \varphi)$ related with original coordinates $(t, r, \theta, \varphi)$ as follows
\begin{eqnarray}\label{compact_hyperboloidal_coordinates}
    t&=&\tau-h(\sigma)\, ,\nonumber\\
    r&=&\frac{r_{\text{h}}}{\sigma}\, ,
\end{eqnarray}
where $h(\sigma)$ is called the height function. The domain of $\sigma$ is restricted to $[0,1]$, where $\sigma=0$ and $\sigma=1$ correspond to the null infinity $\mathscr{I}^+$ and the event horizon $\mathscr{H}$, respectively. To determine the explicit expression of the height function, we consider the asymptotic behavior of the tortoise coordinate. In the following, we always use the unit $M=1$. Based on Eqs. \eqref{rsim1} and \eqref{rsim2} in Appendix \ref{Appendix A}, the singular term of $r_{\star}(\sigma)$, namely $r_{\star 0}$ for $\sigma \rightarrow 0$ and $r_{\star 1}$ for $\sigma \rightarrow 1$, are found out as follows
\begin{eqnarray}\label{rstar01}
    r_{\star 0} &=& \frac{2}{\sigma}-\frac{5\varepsilon(1+\sigma)}{8\sigma}+2\ln\Big(\frac{2}{\sigma}\Big)\, ,\\
    r_{\star 1}&=& 2\ln(1-\sigma)-\frac{1}{4}\varepsilon \ln(1-\sigma)\, .
\end{eqnarray}
Mimicking the in-out and out-in strategies in~\cite{PanossoMacedo:2023qzp}, we obtain the height function
\begin{eqnarray}\label{height_function}
    h(\sigma)=r_{\star 1}-r_{\star 0}\, .
\end{eqnarray}
This ensures that $h\sim -r_{\star}$ when $\sigma\rightarrow 0$, and that $h\sim r_{\star}$ when $\sigma\rightarrow 1$. However, unlike the out-in or in-out strategies, the regular term of $r_{\star}(\sigma)$ is dropped in the height function. This is justified since this term is pure gauge and does not affect the results of QNM spectra. Using the compact hyperboloidal coordinates (\ref{compact_hyperboloidal_coordinates}), Eq. \eqref{tdeq} can be recast into
\begin{eqnarray}\label{maineq}
    \partial_\tau u=Lu\, ,\quad
    L=
    \begin{pmatrix}
        0   & 1   \\
        L_1 & L_2
    \end{pmatrix}\, ,\quad
    u=\begin{pmatrix}
        \phi \\
        \psi
    \end{pmatrix}\, ,
\end{eqnarray}
where $\psi\equiv\partial_{\tau}\phi$. Here we omit the subscript $\ell$ and superscript $\pm$. The differential operators $L_1$ and $L_2$ in Eqs. (\ref{maineq}) are written as
\begin{eqnarray}
    L_{1}&=&\frac{1}{\tilde{w}(\sigma)}\Big[\partial_{\sigma}(p(\sigma)\partial_{\sigma})-q(\sigma)\Big]\, ,\label{operator_L1}\\
    L_{2}&=&\frac{1}{\tilde{w}(\sigma)}\Big[2\gamma(\sigma)\partial_{\sigma}+\partial_{\sigma}\gamma(\sigma)\Big]\, .\label{operator_L2}
\end{eqnarray}
The explicit forms of the above functions are given as follows
\begin{eqnarray}\label{function_p_gamma_w_qx}
    p(\sigma)=\frac{\sigma^2A(\sigma)B(\sigma)}{r_{\text{h}}}\, ,\quad \gamma(\sigma)=h^{\prime}(\sigma)p(\sigma)\, ,\quad \tilde{w}(\sigma)=\frac{c_s^{-2}(\sigma)-\gamma^2(\sigma)}{p(\sigma)}\, ,\quad q(\sigma)=\frac{V(\sigma)}{p(\sigma)}\, ,
\end{eqnarray}
where $\prime$ stands for taking the derivative with respect to $\sigma$. It can be seen that there is a modified term $c_s^2(\sigma)$ in $\tilde{w}(\sigma)$ comparing with the usual one $w(\sigma)=[1-\gamma^2(\sigma)]/p(\sigma)$. Substituting Eqs. (\ref{height_function}), \eqref{Adef} and \eqref{Bdef} into Eqs. \eqref{function_p_gamma_w_qx}, and one can obtain
\begin{eqnarray}
    p(\sigma)&=&\frac{1}{2} (1-\sigma ) \sigma ^2-\frac{1}{32} \sigma ^2 \left(22 \sigma ^7-27 \sigma ^6+10 \sigma -5\right) \varepsilon +\mathcal{O}\left(\varepsilon ^2\right)\, ,\label{explicit_p}\\
    \gamma(\sigma)&=&1-2 \sigma ^2+\frac{1}{16} \left(-44 \sigma ^8+10 \sigma ^7+32 \sigma ^6+5 \sigma ^5+5 \sigma ^4+5 \sigma ^3-13 \sigma ^2\right) \varepsilon +\mathcal{O}(\varepsilon^2)\, ,\label{explicit_gamma}\\
    c_s^{-2}(\sigma)&=&1-9 \left(\sigma ^6-\sigma ^5\right) \varepsilon+\mathcal{O}(\varepsilon^2)\, .\label{velocity}
\end{eqnarray}
Based on these equations, one can confirm that $\tilde{w}$ is regular at the boundary $\sigma=0$ and $\sigma=1$, i.e., $\tilde{w}(0)=(128+52\varepsilon)/(16+5\varepsilon)$ and $\tilde{w}(1)=128(8+2\varepsilon)/(8+\varepsilon)$.
Now, we give a test on the hyperboloidal coordinate \eqref{compact_hyperboloidal_coordinates} having the right causal structure~\cite{PanossoMacedo:2023qzp}. The coordinate transformation \eqref{compact_hyperboloidal_coordinates} yields the following spherical symmetric line element in hyperboloidal coordinates $(\tau,\sigma,\theta,\varphi)$
\begin{eqnarray}
    \mathrm{d}s^2=\frac{1}{\sigma^2}\Big\{A(\sigma)r_{\text{h}}\Big[-p(\sigma)\mathrm{d}\tau^2+2\gamma(\sigma)\mathrm{d}\tau \mathrm{d}\sigma+w(\sigma)\mathrm{d}\sigma^2\Big]+r_{\text{h}}^2(\mathrm{d}\theta^2+\sin^2\theta \mathrm{d}\varphi^2)\Big\}\, .
\end{eqnarray}
After a conformal transformation $\mathrm{d}s^2=\Omega^{-2}\mathrm{d}\Bar{s}^2$ with $\Omega=\sigma$, we get a four-dimensional conformal metric with the line element
\begin{eqnarray}\label{conformal_metric}
    \mathrm{d}\bar{s}^2=\varXi(\sigma)\Big[-p(\sigma)\mathrm{d}\tau^2+2\gamma(\sigma)\mathrm{d}\tau \mathrm{d}\sigma+w(\sigma)\mathrm{d}\sigma^2\Big]+r_{\text{h}}^2(\mathrm{d}\theta^2+\sin^2\theta \mathrm{d}\varphi^2)\, ,
\end{eqnarray}
where $\varXi(\sigma)=A(\sigma)r_{\text{h}}$. The out- and ingoing null vectors are 
\begin{eqnarray}
    \Bar{l}^a=\nu\Big[\Big(\frac{\partial}{\partial\tau}\Big)^a-\frac{1+\gamma}{w}\Big(\frac{\partial}{\partial\sigma}\Big)^a\Big]\, ,\quad
    \Bar{k}^a=\frac{w}{2\varXi\nu}\Big[\Big(\frac{\partial}{\partial\tau}\Big)^a+\frac{1-\gamma}{w}\Big(\frac{\partial}{\partial\sigma}\Big)^a\Big]\, ,
\end{eqnarray}
where $\nu(\sigma)$ is a free Lorentz boost parameter. From Eqs. \eqref{explicit_p} and \eqref{explicit_gamma}, it is found that $\left|\gamma\right|<1$ and $p(\sigma)>0$ for $0< \sigma < 1$ and $0<\varepsilon\ll 1$, which implies $w(\sigma)>0$. At the boundaries, since $p(0)=p(1)=0$ and $\gamma(0)=1\, ,\gamma(1)=-1$, the function $w(\sigma)$ should be evaluated as a limit. These limits yield $w(0)=(128+52\varepsilon)/(16+5\varepsilon)$ and $w(1)=(128+112\varepsilon)/(8+\varepsilon)$. Both values $w(0)$ and $w(1)$ are clearly positive for $0<\varepsilon\ll 1$. In summary, we find that $w(\sigma)>0$ across the entire closed interval $0\le\sigma\le1$, ensuring that the constant-$\tau$ slices are spacelike. By the way, $c_s^{-2}\geqslant1$ ensures that $\tilde{w}(\sigma)\geqslant w(\sigma)$. Accordingly, $\tilde{w}$ remains positive within the whole computational domain.

Finally, we go back to the out- and ingoing null vectors. For $\sigma=0$, they are  
\begin{eqnarray}
    \Bar{l}^a=\nu\Big[\Big(\frac{\partial}{\partial\tau}\Big)^a-\frac{2}{w(0)}\Big(\frac{\partial}{\partial\sigma}\Big)^a\Big]\, ,\quad
    \Bar{k}^a=\frac{w(0)}{2\varXi\nu}\Big(\frac{\partial}{\partial\tau}\Big)^a\, ,
\end{eqnarray}
while for $\sigma=1$, they are
\begin{eqnarray}
       \Bar{l}^a=\nu\Big(\frac{\partial}{\partial\tau}\Big)^a\, ,\quad
    \Bar{k}^a=\frac{w(1)}{2\varXi\nu}\Big[\Big(\frac{\partial}{\partial\tau}\Big)^a+\frac{2}{w(1)}\Big(\frac{\partial}{\partial\sigma}\Big)^a\Big]\, .
\end{eqnarray}
As $\varXi>0$ and $w>0$, the light cones point outward in the boundary of  the conformal spacetime at $\sigma=0$ and $\sigma=1$, so that no boundary information propagates into the domain $\sigma\in[0,1]$. Through the above test, we confirm that our choice of the hyperboloidal coordinate is valid. 

\section{The quasinormal modes of the EFT corrected black hole}\label{QNMs}
Here we present our frequency-domain approach for computing the QNM spectra of the EFT corrected black hole. Considering the Fourier transform of $u(\tau,\sigma)$ with respect to time $\tau$, we have
\begin{eqnarray}
    u(\tau, \sigma)=e^{i \omega t}\tilde{u}(r_\star)=e^{i\omega\tau}u(\sigma)\, ,
\end{eqnarray}
where $u(\sigma)=e^{-i\omega h(\sigma)}\tilde{u}(r_\star)$ from Eqs. (\ref{compact_hyperboloidal_coordinates}). Then the problem of solving QNM spectra turns into an eigenvalue problem $Lu=i\omega u$.

In practice, these frequency-domain calculations will be performed using spectral collocation method. Based on Eqs. \eqref{maineq}, we formulate the problem as an eigenvalue system through spatial discretization. The computational domain is partitioned into a grid $\{\sigma_i\}_{i=0}^N$, where the field variables $\phi$ and $\psi$ are represented by their grid values $\phi_i(\tau)\equiv\phi(\tau,\sigma_i)$ and $\psi_i(\tau)\equiv\psi(\tau,\sigma_i)$. This discretization transforms the original partial differential equations into a coupled system of $(2N+2)$ ordinary differential equations in the temporal variable $\tau$, via
\begin{eqnarray}\label{odes}
    &&\frac{\mathrm{d}}{\mathrm{d}\tau}\begin{bmatrix}
        \phi_i \\
        \psi_i
    \end{bmatrix}
    =\sum_{j=0}^{N}
    \begin{bmatrix}
        0                                                        & \delta_{ij}                       \\
        C_i(\mathbf{D}^2)_{ij}+E_i\mathbf{D}_{ij}+W_i\delta_{ij} & A_i\mathbf{D}_{ij}+B_i\delta_{ij}
    \end{bmatrix}
    \begin{bmatrix}
        \phi_j \\
        \psi_j
    \end{bmatrix}\, ,\quad i=0,1,\cdots,N-1,N\, ,
\end{eqnarray}
where $A_i=A(\sigma_i)$, $B_i=B(\sigma_i)$, $C_i=C(\sigma_i)$, $E_i=E(\sigma_i)$, $W_i=W(\sigma_i)$ and $\delta_{ij}$ is the Kronecker delta. These five functions $C(\sigma)$, $E(\sigma)$, $W(\sigma)$, $A(\sigma)$ and $B(\sigma)$ are directly derived from Eqs. \eqref{operator_L1} Eq. \eqref{operator_L2}. The specific form of the differential matrix $\mathbf{D}$\footnote{We use bold symbols to represent matrices and vectors.} is determined by the chosen Chebyshev-Lobatto grid, given by
\begin{eqnarray}
    \sigma_j=\frac{1}{2}\Big[1+\cos\Big(\frac{j\pi}{N}\Big)\Big]\, ,\quad j=0,1,\cdots,N-1,N\, .
\end{eqnarray}
The expression of the differential matrix $\mathbf{D}$ associated with the Chebyshev-Lobatto grid can be found in~\cite{boyd2001chebyshev,Jaramillo:2020tuu,Cao:2024oud,Trefethen_spectral_method_book}.

To numerically compute the QNM spectra, we discretize the differential operator $L$ using a Chebyshev-Lobatto collocation scheme, which yields its matrix representation $\mathbf{L}$. This discretization effectively reduces the original infinite-dimensional spectral problem to a finite-dimensional algebraic eigenvalue problem
\begin{eqnarray}\label{eigenvalue_porblem}
    \mathbf{L}\mathbf{u}=i\omega \mathbf{u}\, ,
\end{eqnarray}
where the matrix $\mathbf{L}$ corresponds precisely to the coefficient matrix derived from the right-hand side of Eq. \eqref{odes}. While this numerical approach effectively captures physical QNMs, it inevitably generates spurious numerical modes as an artifact of the discretization process. To address this challenge, we implement a robust mode discrimination based on spectral drift analysis, following established methodologies in~\cite{boyd2001chebyshev,Chen:2024mon,BOYD199611,Cownden:2023dam,Cao:2024sot}, which also quantifies mode convergence across grid resolutions. In practice, we use the convention in \cite{Cao:2024sot}, for a mode computed at resolution $N$ we compare it with its nearest counterparts at $N+10$ and $N+20$. A larger drift ratio indicates better cross-grid convergence. Modes with a drift ratio exceeding $10^{3}$ are identified as genuine physical modes.

\begin{figure}[H]
    \centering
    \subfigure[]{\includegraphics[width=0.48\linewidth]{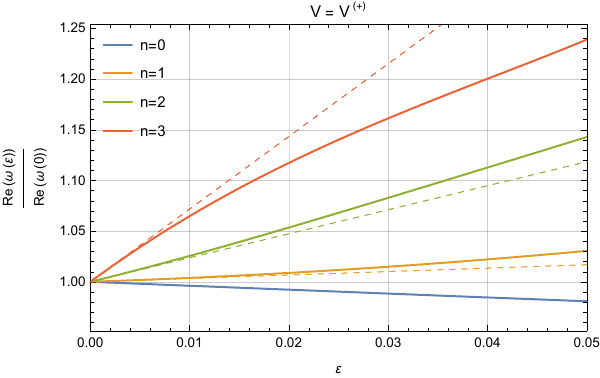}}  \hfill
    \subfigure[]{\includegraphics[width=0.48\linewidth]{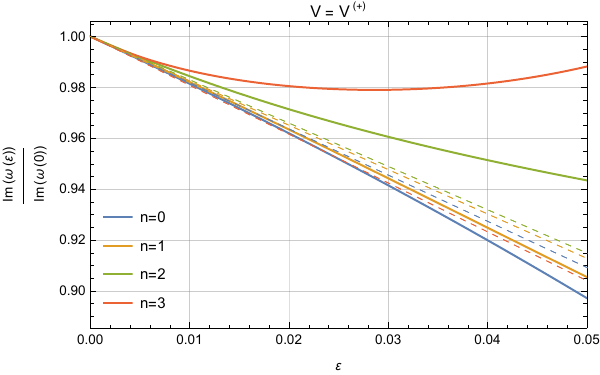}}
    \caption{$\frac{\mathrm{Re}(\omega (\varepsilon))}{\mathrm{Re}(\omega(0))}$ (a) and $\frac{\mathrm{Im}(\omega(\varepsilon))}{\mathrm{Im}(\omega(0))}$ (b) for $\ell=2$ QNM spectra in polar parity perturbations with the overtone number $n=0,1,2,3$. We vary the parameter $\varepsilon$ from $0$ to $0.05$ in increments of $\Delta\varepsilon=0.001$. For each color, the dashed line is tangent to the solid line at $\varepsilon=0$. The calculation is performed on the Chebyshev-Lobatto grid with $N=200$.}
    \label{fig:overtonep}
\end{figure}

\begin{figure}[H]
    \centering
    \subfigure[]{\includegraphics[width=0.48\linewidth]{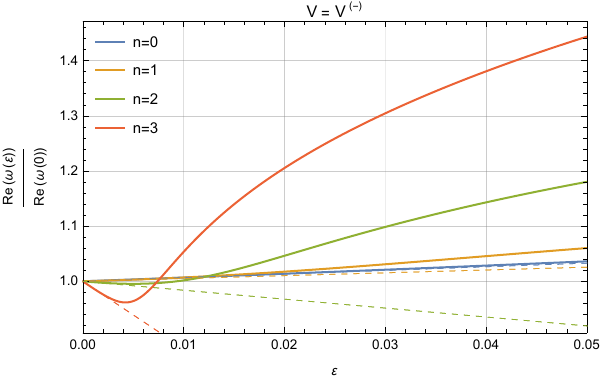}}  \hfill
    \subfigure[]{\includegraphics[width=0.48\linewidth]{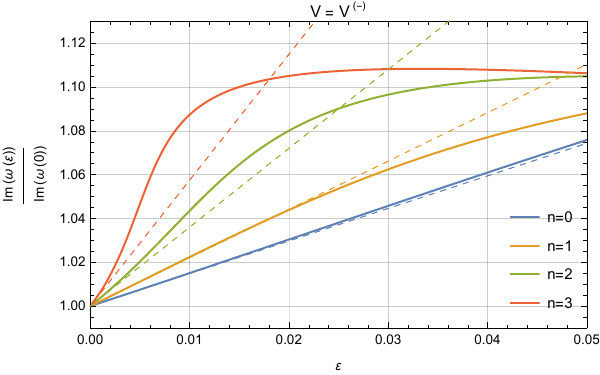}}
    \caption{$\frac{\mathrm{Re}(\omega (\varepsilon))}{\mathrm{Re}(\omega(0))}$ (a) and $\frac{\mathrm{Im}(\omega(\varepsilon))}{\mathrm{Im}(\omega(0))}$ (b) for $\ell=2$ QNM spectra in axial parity perturbations with the overtone number $n=0,1,2,3$. We vary the parameter $\varepsilon$ from $0$ to $0.05$ in increments of $\Delta\varepsilon=0.001$. For each color, the dashed line is tangent to the solid line at $\varepsilon=0$. The calculation is performed on the Chebyshev-Lobatto grid with $N=200$.}
    \label{fig:overtonem}
\end{figure}

Note that the upper bound of $\varepsilon$ is $\varepsilon_\text{lim}=3$ (see the discussion following its introduction in Sec. \ref{theory_solution_perturbation}), so we can concentrate on the QNMs with $\ell=2$ for $\varepsilon$ varying from $0$ to $0.05$ in increments of $\Delta\epsilon=0.001$. The resulting spectra are displayed in Figs. \ref{fig:overtonep} and \ref{fig:overtonem} by plotting $\frac{\mathrm{Re}(\omega (\varepsilon))}{\mathrm{Re}(\omega(0))}$ and $\frac{\mathrm{Im}(\omega(\varepsilon))}{\mathrm{Im}(\omega(0))}$, where $\omega(0)$ refers to the spectrum of the Schwarzschild black hole. The QNM spectra derived via our hyperboloidal approach exhibit concordance with those obtained using the phase-amplitude method~\cite{Silva:2024ffz}. The tiny  discrepancies at $\mathcal{O}(\varepsilon^2)$ originate in the divergent $\mathcal{O}(\varepsilon)$ approximation schemes employed by each methodology. For the Schwarzschild black hole $(\varepsilon=0)$, the QNMs of polar and axial perturbations are identical. This phenomenon is known as isospectrality of GR.
However, as $\varepsilon$ increases from $0$, as can be seen by comparing Figs. \ref{fig:overtonep} and \ref{fig:overtonem}, the isospectrality is broken. For the fundamental modes, the real and imaginary parts of the polar parity perturbations increase and that of the axial parity decrease as $\varepsilon$ increases. Furthermore, the solid lines and their tangent lines show good alignment. For higher overtones, the real and imaginary parts of both polar and axial parity perturbations exhibit more complex behavior as $\varepsilon$ increases, and the differences between the solid lines and their tangent lines become larger. Moreover, the absolute value of the slope is larger for higher overtones, except for the right panel of Fig. \ref{fig:overtonep}. Thus, higher overtones are more sensitive to $\varepsilon$, which is in agreement with previous work~\cite{Silva:2024ffz}.

The sensitivity of $\omega(\varepsilon)$ is captured by its linear term, which can be determined not only by directly calculating the slopes of tangent lines at $\varepsilon=0$, but also through the perturbation theory, a standard approach in quantum mechanics. Let us consider a linear operator $A$ act on a complex vector space with inner product $\langle \cdot,\cdot\rangle$, and denote its adjoint as $A^{\dagger}$, which is defined as $\langle A^{\dagger}u,v\rangle=\langle u,Av\rangle$. The left $u_i$ and right $v_i$ eigenvectors are characterized, respectively, as 
\begin{eqnarray}
    A^{\dagger}u_i=\Bar{\lambda}_iu_i\, ,\quad Av_i=\lambda_iv_i\, ,
\end{eqnarray}
with $\Bar{\lambda}_i$ as the complex conjugate of $\lambda_i$. For $A(\epsilon)$ as an operator depending on a small parameter $\epsilon>0$, we consider a linear perturbation of $A$ by a bounded operator $\delta A$,
\begin{eqnarray}
    A(\epsilon)=A+\epsilon \delta A+\mathcal{O}(\epsilon^2)\, ,\quad \lVert \delta A\rVert<\infty \, .
\end{eqnarray}
Consider the eigenvalue problem associated with $A(\epsilon)$,
\begin{eqnarray}  A(\epsilon)v_i(\epsilon)=\lambda_i(\epsilon)v_i(\epsilon)\, .
\end{eqnarray}
If all of the eigenvalues $\lambda_i$ are simple, then according to the perturbation theory, the expansions of eigenvalues and eigenvectors are, respectively,
\begin{eqnarray}
    \lambda_i(\epsilon)&=&\lambda_i(0)+\epsilon\lambda_i^{(1)}+\mathcal{O}(\epsilon^2)\, ,\\
    v_i(\epsilon)&=&v_i(0)+\epsilon v_i^{(1)}+\mathcal{O}(\epsilon^2)\, ,
\end{eqnarray}
with
\begin{eqnarray}
    \lambda_i^{(1)}&=&\frac{\langle u_i,\delta Av_i\rangle}{\langle u_i,v_i \rangle}\, ,\label{lambda_perturbation}\\
    v_i^{(1)}&=&\sum_{j\ne i}{\frac{\langle u_j,\delta Av_i\rangle}{(\lambda_i-\lambda_j)\langle u_j,v_j \rangle}v_i}\label{vi_perturbation}\, ,
\end{eqnarray}
where we have used the fact that $\langle u_i,v_j \rangle=0$ for $i\neq j$. The coefficient $\lambda_i^{(1)}$ can characterize the sensitivity of an eigenvalue as the small parameter $\epsilon$ changing from $0$. In numerical calculation, we use matrices to approximate operators, and the inner product is chosen as the Euclidean inner product $\langle \cdot,\cdot\rangle_2$. We note that Eq. \eqref{lambda_perturbation} is independent of the choice of the inner product, and we will prove it for the matrix version as follows:

The new inner product between vector $\mathbf{u}$ and $\mathbf{v}$ can be derived from the old one as $\langle \mathbf{u}, \mathbf{v} \rangle_{\text{G}}=\langle \mathbf{u},\mathbf{G} \cdot\mathbf{v} \rangle_2$, so the new adjoint matrix $\hat{\mathbf{A}}$ of $\mathbf{A}$ has a relation between the old one, i.e., $\hat{\mathbf{A}}=\mathbf{G}^{-1}\cdot\mathbf{A}^{\star}\cdot\mathbf{G}$, where $\mathbf{G}$ is a positive definite Hermitian Gram matrix and $\star$ denotes the Hermitian conjugate. Then, $\mathbf{A}$'s new right eigenvectors are identical to the original one, and new left eigenvectors $\tilde{\mathbf{u}}_i$ are
 \begin{eqnarray}
     \tilde{\mathbf{u}}_i=\mathbf{G}^{-1}\cdot\mathbf{u}_i\, .
 \end{eqnarray}
Given an arbitrary vector $\mathbf{w}$, the new inner product of $\mathbf{u}_i$ and $\mathbf{w}$ is
\begin{eqnarray}\label{u_w_G_and_u_w_2}
\langle \tilde{\mathbf{u}}_i,\mathbf{w}\rangle_{\text{G}}=\langle \tilde{\mathbf{u}}_i,\mathbf{G}\cdot\mathbf{w}\rangle_2=\langle \mathbf{G}^{-1}\cdot\mathbf{u}_i,\mathbf{G}\cdot\mathbf{w}\rangle_2=\langle \mathbf{u}_i,(\mathbf{G}^{-1})^{\star}\cdot\mathbf{G}\cdot\mathbf{w}\rangle_2=\langle \mathbf{u}_i,\mathbf{w}\rangle_2\,.
\end{eqnarray}
Denoting $\tilde{\lambda}_i^{(1)}$ as the coefficients calculated by the new inner product, from Eq. (\ref{u_w_G_and_u_w_2}), $\lambda_i^{(1)}$ is unchanged, namely
\begin{eqnarray}
    \tilde{\lambda}_i^{(1)}&=&\frac{\langle \tilde{\mathbf{u}}_i,\delta \mathbf{A}\cdot\mathbf{v}_i\rangle_{\text{G}}}{\langle \tilde{\mathbf{u}}_i,\mathbf{v}_i \rangle_{\text{G}}}=\frac{\langle \mathbf{u}_i,\delta \mathbf{A}\cdot \mathbf{v}_i\rangle_2}{\langle \mathbf{u}_i,\mathbf{v}_i \rangle_2}=\lambda_i^{(1)}\, .\label{new_lambda_perturbation}
\end{eqnarray}

Now, coming back to our present case, we use the formula \eqref{lambda_perturbation} with  $A(\varepsilon)=L(\varepsilon)/i\, ,\epsilon=\varepsilon$ and $\delta A=L^{\prime}(\varepsilon)/i|_{\varepsilon=0}$ to obtain the linear term $\omega_n^{(1)}$ of $\omega_n(\varepsilon)$. The real part and the imaginary part of $\omega_n^{(1)}$ correspond to the slopes of the tangent lines of $\frac{\mathrm{Re}(\omega (\varepsilon))}{\mathrm{Re}(\omega(0))}$ and $\frac{\mathrm{Im}(\omega(\varepsilon))}{\mathrm{Im}(\omega(0))}$ at $\varepsilon=0$, respectively, i.e.,
\begin{eqnarray}\label{k_Re_Im}
    k^{(\mathrm{Re})}_n=\frac{\mathrm{Re}(\omega_n^{(1)})}{\mathrm{Re}(\omega_n(0))}\, ,\quad k^{(\mathrm{Im})}_n=\frac{\mathrm{Im}(\omega_n^{(1)})}{\mathrm{Im}(\omega_n(0))}\, .
\end{eqnarray}
Using Eqs. (\ref{k_Re_Im}), we show the results in Tables \ref{tab:Vp_kReIm} and \ref{tab:Vm_kReIm}. Our calculation results of $\omega_n(0)$ are highly consistent with the high-precision numerical results obtained by Leaver (1985)~\cite{Leaver:1985ax} using the continuous fraction method. The slopes calculated through perturbation method ($k^{(\mathrm{Re}/\mathrm{Im})}_n$) are close to those calculated by interpolating the QNM data ($\hat{k}^{(\mathrm{Re}/\mathrm{Im})}_n$), and the deviation becomes greater as $n$ increases. 
\begin{table}[H]
    \centering
    \begin{tabular}{cccccc}
    \hline
        $n$ & $\omega_n(0)$ & $k^{(\mathrm{Re})}_n$& $\hat{k}^{(\mathrm{Re})}_n$ &  $k^{(\mathrm{Im})}_n$ & $\hat{k}^{(\mathrm{Im})}_n$\\
        \hline
        $0$ & $0.37367 + 0.08896i$ & $-0.3856$ &$-0.3856$ &  $-1.8162$&$-1.8162$ \\
        $1$ & $0.34671 + 0.27392i$ & $ 0.3362$& $0.3362$ &   $-1.7442$&$-1.7442$\\
        $2$ & $0.30105 + 0.47828i$ & $2.3689$& $2.3689$  &   $-1.7024$&$-1.7024$\\
        $3$ & $0.25150 + 0.70515i$ &  $7.1685$& $ 7.1690$ &   $-1.9200$&$-1.9198$\\
        \hline
    \end{tabular}
    \caption{$\omega_n(0), k^{(\mathrm{Re,Im})}_n$ and $\hat{k}^{(\mathrm{Re,Im})}_n$ of $\ell=2$ polar parity QNMs with the overtone number $n=0,1,2,3$ are shown. $k^{(\mathrm{Re,Im})}_n$ are obtained from \eqref{k_Re_Im}, while $\hat{k}^{(\mathrm{Re,Im})}_n$ are the slopes of the tangent lines in Fig. \ref{fig:overtonep}. The calculation is performed on the Chebyshev-Lobatto grid with $N=200$.}
    \label{tab:Vp_kReIm}
\end{table}

\begin{table}[H]
    \centering
    \begin{tabular}{cccccc}
    \hline
        $n$ & $\omega_n(0)$ & $k^{(\mathrm{Re})}_n$& $\hat{k}^{(\mathrm{Re})}_n$ &  $k^{(\mathrm{Im})}_n$ & $\hat{k}^{(\mathrm{Im})}_n$ \\
        \hline
        $0$ & $0.37367 + 0.08896i$ & $0.6584 $& $0.6585$ &  $1.4852$&$1.4853$ \\
        $1$ & $0.34671 + 0.27392i$ & $ 0.5113$& $0.5113$  &$2.2054$&$2.2054$\\
        $2$ & $0.30105 + 0.47828i$ & $-1.6186$ &  $-1.6179$ &$3.5993$&$3.5985$\\
        $3$ & $0.25150 + 0.70515i$ &  $-12.2788$ & $-12.1647$ & $5.7396$&$5.7400$\\
        \hline
    \end{tabular}
    \caption{$\omega_n(0), k^{(\mathrm{Re,Im})}_n$ and $\hat{k}^{(\mathrm{Re,Im})}_n$ of $\ell=2$ axial parity QNMs with the overtone number $n=0,1,2,3$ are shown. $k^{(\mathrm{Re,Im})}_n$ are obtained from \eqref{k_Re_Im}, while $\hat{k}^{(\mathrm{Re,Im})}_n$ are the slopes of the tangent lines in Fig. \ref{fig:overtonem}. The calculation is performed on the Chebyshev-Lobatto grid with $N=200$.}
    \label{tab:Vm_kReIm}
\end{table}

\section{Time-domain waveform}\label{TD_wave_form}
In this section we perform the time-domain analysis to study the affect of $\varepsilon$ on the waveform. In order to numerically solve the ordinary differential equations (\ref{odes}) derived from (\ref{maineq}), we employ a discrete time evolution scheme based on a sixth-order Hermite integration method~\cite{markakis2019timesymmetry,Markakis:2023pfh, DaSilva:2024yea,OBoyle:2022lek,OBoyle:2022yhp}, which is inherently implicit. Choosing a fixed time step $\Delta\tau$, we obtain
\begin{eqnarray}
    \mathbf{u}\left((i+1)\Delta\tau\right) = \mathbf{U}\cdot\mathbf{u}(i\Delta\tau),\quad i=0,1,\cdots\, ,
\end{eqnarray}
where $\mathbf{U}$ is called the evolution matrix whose explicit expression is
\begin{eqnarray}
    \mathbf{U}=\mathbf{I}+(\Delta \tau \mathbf{L})\cdot\Big(\mathbf{I}+\frac{1}{60}(\Delta \tau \mathbf{L})\cdot(\Delta \tau \mathbf{L})\Big)\cdot\left(\mathbf{I}-\frac{\Delta \tau}2\mathbf{L}\cdot\left(\mathbf{I}-\frac{\Delta \tau}5\cdot\left(\mathbf{I}-\frac{\Delta \tau}{12}\mathbf{L}\right)\right)\right)^{-1},
\end{eqnarray}
and $\mathbf{I}$ denotes the identity matrix with dimensions matching those of $\mathbf{L}$. Since this scheme is unconditionally stable, the time step $\Delta\tau$ is not constrained by the Courant condition. Furthermore, $\Delta\tau$ need not remain constant and can instead vary with the step index $i$. For simplicity, however, we adopt uniform time steps in our implementation.

\begin{figure}[H]
    \centering
    \subfigure[]{\includegraphics[width=0.48\linewidth]{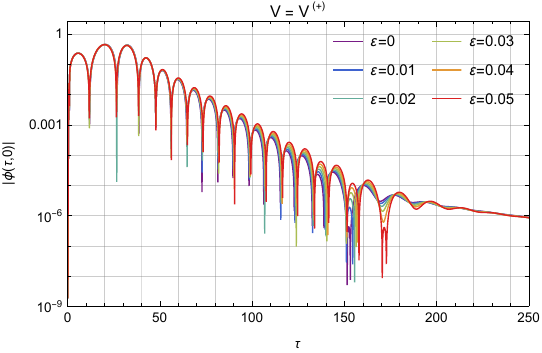}\label{Vptimedomaina}}  \hfill
    \subfigure[]{\includegraphics[width=0.48\linewidth]{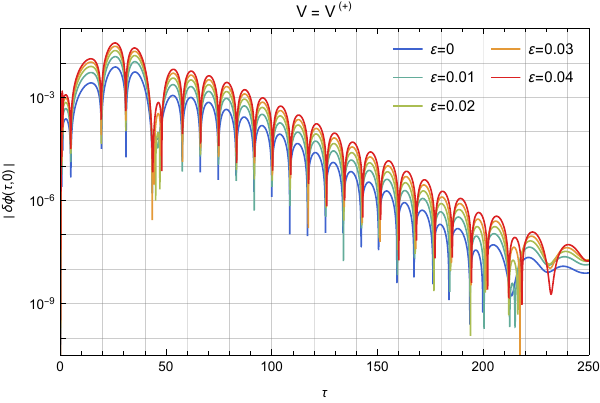}\label{Vptimedomainb}}
    \caption{Time-domain waveform at $\sigma=0$ for polar parity perturbation. (a) The absolute value of $\phi(\tau,0)$ for $\ell=2$ and $\varepsilon$ varying from $0$ to $0.05$ in increments of $\Delta\varepsilon=0.01$, together with $\tau$ varies from $0$ to $250$. (b) The absolute value of the difference between $\phi(\tau,0)$ in the case $\varepsilon=0.01\sim 0.05$ and $\varepsilon=0$. The calculation is performed on the Chebyshev-Lobatto grid with $N=100$.}
    \label{Vptimedomain}
\end{figure}

\begin{figure}[H]
    \centering
    \subfigure[]{\includegraphics[width=0.48\linewidth]{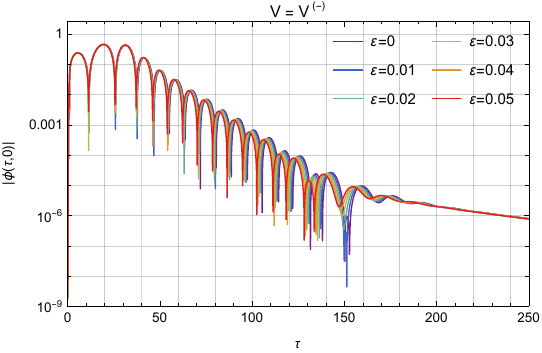}\label{Vmtimedomaina}}  \hfill
    \subfigure[]{\includegraphics[width=0.48\linewidth]{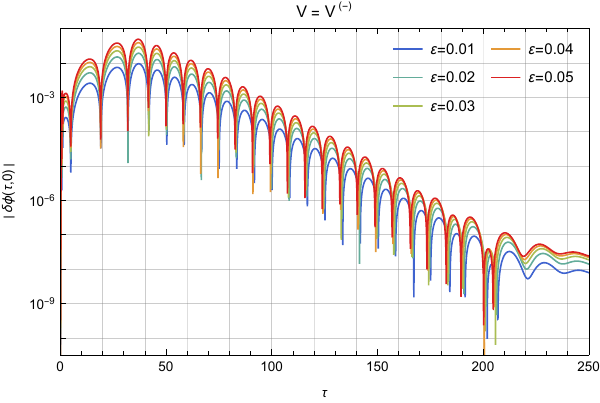}\label{Vmtimedomainb}}
    \caption{Time-domain waveform at $\sigma=0$ for axial parity perturbation. (a) The absolute value of $\phi(\tau,0)$ for $\ell=2$ and $\varepsilon$ varying from $0$ to $0.05$ in increments of $\Delta\varepsilon=0.01$, together with $\tau$ varying from $0$ to $250$. (b) The absolute value of the difference between $\phi(\tau,0)$ in the case $\varepsilon=0.01\sim 0.05$ and $\varepsilon=0$. The calculation is performed on the Chebyshev-Lobatto grid with $N=100$.}
    \label{Vmtimedomain}
\end{figure}

As for the initial data, we impose a Gaussian pulse in the $\sigma$ coordinate, i.e.,
\begin{eqnarray}
    \phi(\tau=0,\sigma)&=&a_0\exp\bigg[-\frac{(\sigma-c_0)^2}{2b_0^2}\bigg]\, ,\nonumber\\
    \psi(\tau=0,\sigma)&=&0\, ,
    \label{initial condition}
\end{eqnarray}
where $a_0$ is the amplitude, $b_0$ is the width and $c_0$ is the position of the Gaussian pulse. The time-domain waveforms of polar and axial parity perturbations for $\varepsilon$ ranging from $0$ to $0.05$ with spacing $0.01$ are shown in Figs. \ref{Vptimedomain} and \ref{Vmtimedomain}, where the observer is located at infinity and the initial data are chosen as (\ref{initial condition}) with $a_0=1$, $b_0=1/(10\sqrt{10})$ and $c_0=1/5$.

The time-domain waveform in Figs. \ref{Vptimedomain} and \ref{Vmtimedomain} can be divided into three stages: the prompt response within about $\tau=0\sim20$, the ringdown within about $\tau=20\sim200$ and the tail within about $\tau=200\sim250$. In the prompt response stage, the amplitude of the wave function $\phi$ increases. In the ringdown stage, $\phi$ undergoes a damped oscillation. In the tail stage, $\phi$ is in power-law decay. As can be seen from the Figs. \ref{Vptimedomaina} and \ref{Vmtimedomaina}, the relative difference of the waveform is non-negligible in the ringdown stage, while it is much smaller in the prompt response stage and the tail stage. Based on Figs. \ref{Vptimedomainb} and \ref{Vmtimedomainb}, we can roughly conclude that  $\delta\phi$ is linearly dependent on $\varepsilon$.

\begin{figure}[H]
    \centering
    \includegraphics[width=0.6\linewidth]{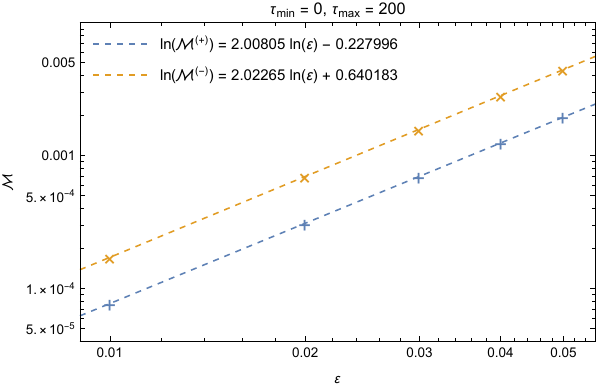}
     \caption{The mismatch of the polar parity $(+)$ and axial parity $(-)$ time-domain waveform between $\varepsilon=0$ and $\varepsilon=0.01$, $0.02$, $0.03$, $0.04$, $0.05$. We set $\tau_{\text{min}}=0$ and $\tau_{\text{max}}=200$. The figure is plotted using double logarithmic coordinates, and the dashed lines are the fitting lines. The calculation is performed on the Chebyshev-Lobatto grid with $N=100$.}
    \label{mismatch}
\end{figure}

Apart from subtracting directly, another way to characterize the difference between two waveforms is to calculate the mismatch, which is defined as
\begin{eqnarray}
    \mathcal{M}\left[h_1(\tau),h_2(\tau)\right] \equiv 1- \frac{\langle h_1(\tau), h_2(\tau)\rangle}{\sqrt{\langle h_1(\tau), h_1(\tau)\rangle\langle h_2(\tau), h_2(\tau)\rangle}}\, ,
    \label{def_mismatch}
\end{eqnarray}
where the inner product above is
\begin{eqnarray}
    \langle h_1(\tau), h_2(\tau)\rangle \equiv \int^{\tau_{\text{max}}}_{\tau_{\text{min}}}{\overline{h_1(\tau)}h_2(\tau)\mathrm{d}\tau}\, .
\end{eqnarray}
Then in Fig. \ref{mismatch} we show the mismatch of the waveform between $\varepsilon=0$ and $\varepsilon=0.01$, $0.02$, $0.03$, $0.04$, $0.05$, where we have set $\tau_{\text{min}}=0$ and $\tau_{\text{max}}=200$ . We find that the slope of the fitting line is about $2$, so $\mathcal{M}\propto \varepsilon^2$. Although the QNM spectra change a lot when $\varepsilon$ varies from $0$ to $0.05$, the time-domain waveform changes a little. Moreover, Fig. \ref{mismatch} implies that the change of $\varepsilon$ from $0$ to $0.05$ results in a continuous rather than abrupt change in the time-domain waveform.

We give an explanation on the result $\mathcal{M}\propto \varepsilon^2$. Since $u(\tau)= \mathrm{e}^{L\tau}u(0)$ and $L=L^{(0)}+\varepsilon L^{(1)}+\mathcal{O}(\varepsilon^2)$, the waveforms for small $\varepsilon$ satisfy
\begin{eqnarray}
    h(\tau, \varepsilon)=h^{(0)}(\tau)+\varepsilon h^{(1)}(\tau)+\mathcal{O}(\varepsilon^2)\,, 
    \label{h_varepsilon}
\end{eqnarray}
where $h^{(0)}(\tau)=h(\tau, 0)$ and $h_{1}(\tau)$ is the leading-order correction. We substitute \eqref{h_varepsilon} into the definition of mismatch \eqref{def_mismatch} and retain up to $\mathcal{O}(\varepsilon^2)$, then we can obtain
\begin{eqnarray}
    \mathcal{M}=\frac{\langle h^{(0)},h^{(1)}\rangle^2-\langle h^{(0)},h^{(0)}\rangle\langle h^{(1)},h^{(1)}\rangle}{2\langle h^{(0)},h^{(0)}\rangle^2}\varepsilon^2+\mathcal{O}(\varepsilon^3)\, .
\end{eqnarray}
Here we have used the fact that these waveforms are real. Therefore, we have confirmed that the mismatch $\mathcal{M}$ is proportional to $\varepsilon^2$.

\section{Pseudospectrum}\label{stability_analysis}
In this section, we analyze the stability of QNM spectrum of certain fixed $\varepsilon$ under  environmental effects. Pseudospectrum, which is a powerful approach to study the spectrum instability~\cite{Destounis:2023ruj, Jaramillo:2020tuu,trefethen2020spectra}, is adopted. Given $\epsilon>0$, and a norm $\lVert\cdot\rVert$, the $\epsilon$-pseudospectrum for an operator $A$ is defined as~\cite{trefethen2020spectra}
\begin{eqnarray}\label{def of pseudo 1}
    \sigma_\epsilon(A)=\{\omega\in\mathbb{C}:\lVert(A-\omega I)^{-1}\rVert>\epsilon^{-1}\}\,,
\end{eqnarray}
where $I$ is the identity operator, and $R_{A}(\omega)=(A-\omega I)^{-1}$ is called the resolvent operator. This definition is the most suitable for computing the pseudospectrum. Another equivalent and more intuitive definition of the pseudospectrum is~\cite{trefethen2020spectra}
\begin{eqnarray}\label{def of pseudo 2}
    \sigma_{\epsilon}(A) =\{\omega\in\mathbb{C}: \exists\delta A \, ,\text{with}\, \lVert\delta A\rVert<\epsilon\, ,\text{such that}\, \omega\in\sigma(A+\delta A)\}\, .
\end{eqnarray}

The definition (\ref{def of pseudo 2}) illustrates how the pseudospectrum characterizes the (in)stability: Given a small perturbation $\delta A$ whose norm is smaller than $\epsilon$, the pseudospectrum contains all possible spectra of $A+\delta A$. In the limit $\epsilon\to0$, the pseudospectrum set $\sigma_\epsilon(A)$ converges to the spectrum $\sigma(A)$. The parameter $\epsilon$ quantifies the size of the perturbation $\delta A$, providing a rigorous characterization for operator perturbations. The boundary of $\sigma_\epsilon(A)$ characterizes ``how far" the spectrum can reach by adding a perturbation $\delta A$ with a norm smaller than $\epsilon$. For black holes, theoretically we usually assume they are isolated and then derive the perturbation equation. However, the Universe is filled with various forms of matter, such as interstellar clouds, gas, dust and dark matter, and thus the perturbation equation is modified by these forms of matter. In order to study the (in)stability of QNM spectra, we need to characterize the magnitude of the change of the perturbation equation and then analyze the change of the spectra in such magnitude. Particularly, within hyperboloidal framework, we need to characterize the magnitude of the change $\delta L$ of $L$ in the perturbation equation \eqref{maineq}. Fortunately, the energy norm (determined after) is a good choice, and the pseudospectrum $\sigma_{\epsilon}(L/i)$ can be used to study the (in)stability of QNM spectra. By definition \eqref{def of pseudo 2}, $\sigma_{\epsilon}(L/i)$ means the set of the QNM spectra $\omega\in \sigma((L+\delta L)/i)$ due to all possible changes of the perturbation equation with $||\delta L/i||<\epsilon$. In other words, $\epsilon$ characterizes the bound of the change of the perturbation equation, and $\sigma_{\epsilon}(L/i)$ is the reachable range of the QNM spectra under such bound. For fixed $\epsilon$, the closer the boundary of $\sigma_{\epsilon}(L/i)$ lies to a specific $\omega_n$, the greater the stability of such mode $\omega_n$.

\subsection{Energy inner product and energy norm}\label{energynorm}
\begin{figure}[H]
    \centering
    \subfigure[]{\includegraphics[width=0.28\linewidth]{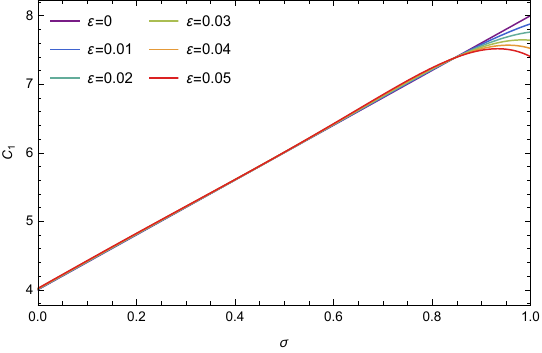}}  \hfill
    \subfigure[]{\includegraphics[width=0.3\linewidth]{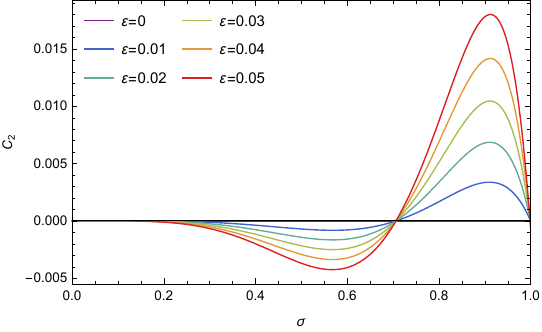}} \hfill
    \subfigure[]{\includegraphics[width=0.28\linewidth]{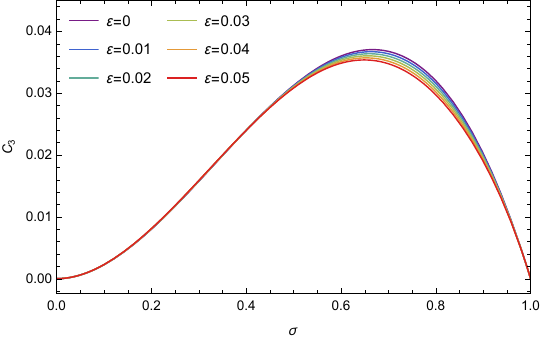}} \\
    \subfigure[]{\includegraphics[width=0.28\linewidth]{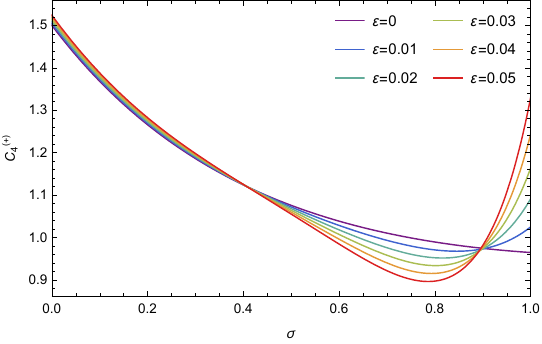}}  \hspace{1.2cm}
    \subfigure[]{\includegraphics[width=0.28\linewidth]{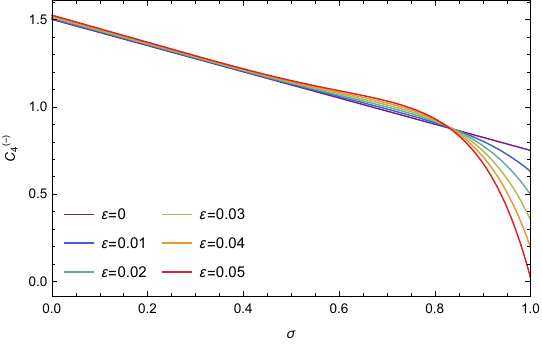}}

    \caption{Coefficients (a) $C_1$, (b) $C_2$, (c) $C_3$ and (d),(e) $C_4$ for polar $(+)$ and axial $(-)$ perturbations.}
    \label{fig:Coefficients}
\end{figure}

Both the shape and the structure of the pseudospectrum are strongly influenced by the choice of the norm. Therefore, before calculating the pseudospectrum, we need to designate a physical relevance norm. We choose the energy norm which is motivated by~\cite{Jaramillo:2020tuu, Gasperin:2021kfv, Besson:2024adi,Cai:2025irl}. Usually, one can find that the wave functions $\phi$ can be viewed as propagating in an effective $(1+1)$-dimensional Minkowski spacetime with an effective potential. Then one can define the energy inner product and energy norm through the effective energy in $\tau=\text{constant}$ hypersurface $\varSigma_\tau$. However, the perturbation equation under consideration incorporates a sound velocity, therefore the process is no longer so simple. The master perturbation equation \eqref{tdeq} in the time domain can still be obtained from the following equation of motion of a complex scalar field:
\begin{eqnarray}\label{eqom}
    (\Box-V)\phi=0\, ,
\end{eqnarray}
where $\Box$ is the d'Alembert operator in an effective $1+1$-dimensional spacetime (no longer the Minkowski spacetime) with an effective metric
\begin{eqnarray}
    \tilde{g}_{ab}\mathrm{d}x^{a}\mathrm{d}x^b=-c_s^2\left(r_{\star}\right)\mathrm{d}t^2+\mathrm{d}r_{\star}^2\, .
\end{eqnarray}
Equation \eqref{eqom} can be obtained from the following Lagrangian
\begin{eqnarray}
    \mathscr{L} =-\frac{1}{4}\tilde{g}^{ab}\Big(\nabla_a\phi\nabla_b\Bar{\phi}+\nabla_a\Bar{\phi}\nabla_b\phi\Big)-\frac{1}{2}V\Bar{\phi}\phi\, ,
\end{eqnarray}
where $\nabla_a$ is the adaptive derivative with respect to the metric $\tilde{g}_{ab}$. From this Lagrangian we can get the effective energy-momentum tensor 
\begin{eqnarray}
    \tilde{T}_{ab}=\frac{1}{2}\Big[\nabla_a\Bar{\phi}\nabla_b\phi-\frac{1}{2}\tilde{g}_{ab}\Big(\tilde{g}^{cd}\nabla_c\Bar{\phi}\nabla_d\phi+V\phi\Bar{\phi}\Big)+\text{c.c}.\Big]\, ,
\end{eqnarray}
where c.c. refers to the complex conjugate term. Then we can obtain the effective energy on a spacelike constant-$\tau$ hypersurface $\varSigma_\tau$ as
\begin{eqnarray}\label{Edef}
    \tilde{E}=\int_{\varSigma_{\tau}}{\tilde{T}_{ab}t^an^b\mathrm{d}}\varSigma _{\tau}=\int_0^1{J_an^a\sqrt{A^2B}\sqrt{{g^{\prime}}^2-{h^{\prime}}^2}\mathrm{d}\sigma}\, ,
\end{eqnarray}
in which $g^{\prime}(\sigma)=-1/p(\sigma)$, $t^a$ is a timelike Killing vector, $n^a$ is a future-pointing timelike unit vector normal to $\varSigma_{\tau}$, and their expressions are, 
\begin{eqnarray}
    t^a=\Big(\frac{\partial}{\partial \tau}\Big)^a\, ,\quad
    n^a=\frac{1}{\sqrt{A^2B}\sqrt{{g^{\prime}}^2-{h^{\prime}}^2}}\Big[w\Big(\frac{\partial}{\partial \tau}\Big)^a-\gamma\Big(\frac{\partial}{\partial\sigma}\Big)^a\Big]\, .
\end{eqnarray}
In Eq. (\ref{Edef}), the integrand $J_a n^a$ is
\begin{eqnarray}\label{Jana}
    J_a n^a\equiv \tilde{T}_{ab}t^a n^b &=& \frac{1}{\sqrt{A^2B}\sqrt{{g^{\prime}}^2-{h^{\prime}}^2}}(w\tilde{T}_{\tau\tau}-\gamma\tilde{T}_{\tau\sigma})\nonumber\\
     &=&\frac{1}{\sqrt{A^2B}\sqrt{{g^{\prime}}^2-{h^{\prime}}^2}}\Big[-\frac{{g^{\prime}}^2-{h^{\prime}}^2}{g^{\prime}}\tilde{T}_{tt}+\frac{h^{\prime}}{g^{\prime}}(-h^{\prime}\tilde{T}_{tt}+g^{\prime}\tilde{T}_{t r_{\star}})\Big]\nonumber\\
   &=&\frac{1}{\sqrt{A^2B}\sqrt{{g^{\prime}}^2-{h^{\prime}}^2}}(-g^{\prime}\tilde{T}_{tt}+h^{\prime}\tilde{T}_{t r_{\star}})\, ,
\end{eqnarray}
with
\begin{eqnarray}
    \tilde{T}_{tt} &=& \frac{1}{2}\partial_{t}\bar{\phi}\partial_t\phi + \frac{c_s^2}{2}\partial_{r_{\star}}\bar{\phi}\partial_{r_{\star}}\phi + \frac{c_s^2}{2}V \phi\bar{\phi} \nonumber\\
    &=& \frac{1}{2} \Big(1+\frac{{h^{\prime}}^2 c_s^2}{{g^{\prime}}^2}\Big)\partial_{\tau}\bar{\phi}\partial_{\tau}\phi+\frac{h^{\prime}c_s^2}{2{g^{\prime}}^2}(\partial_{\tau}\bar{\phi}\partial_{\sigma}\phi+\partial_{\tau}\phi\partial_{\sigma}\bar{\phi})+\frac{c_s^2}{2{g^{\prime}}^2}\partial_{\sigma}\bar{\phi}\partial{\sigma}\phi+\frac{c_s^2}{2}V\phi\bar{\phi}\, ,\label{Ttt}\\
    \tilde{T}_{t r_{\star}}&=&\frac{1}{2}(\partial_{t}\bar{\phi}\partial_{r_{\star}}\phi+\partial_t\phi\partial_{r_{\star}}\bar{\phi})\nonumber\\
    &=&\frac{h^{\prime}}{g^{\prime}}\partial_{\tau}\phi\partial_{\tau}\bar{\phi}+\frac{1}{2g^{\prime}}(\partial_{\tau}\bar{\phi}\partial_{\sigma}\phi+\partial_{\tau}\phi\partial_{\sigma}\bar{\phi})\, .\label{Ttrstar}
\end{eqnarray}
Substituting Eqs. \eqref{Jana}-\eqref{Ttt} and Eq. \eqref{Ttrstar} into Eq. \eqref{Edef}, we can get
\begin{eqnarray}
    \tilde{E}=\int_0^1{\mathrm{d}\sigma}\Big[C_1\partial_{\tau}\bar{\phi}\partial_{\tau}\phi+C_2(\partial_{\tau}\bar{\phi}\partial_{\sigma}\phi+\partial_{\tau}\phi\partial_{\sigma}\bar{\phi})+C_3\partial_{\sigma}\bar{\phi}\partial_{\sigma}\phi+C_4\bar{\phi}\phi\Big]\, ,
\end{eqnarray}
where the coefficients are
\begin{eqnarray}
    C_1=-\frac{{g^{\prime}}^2+{h^{\prime}}^2(c_s^2-2)}{2g^{\prime}}\,,\quad
    C_2=\frac{h^{\prime}(1-c_s^2)}{2g^{\prime}}\,,\quad
    C_3=-\frac{c_s^2}{2g^{\prime}}\,,\quad
    C_4=-\frac{g^{\prime}c_s^2 V}{2}\,,
\end{eqnarray}
and in our calculations, we truncate these coefficients to $\mathcal{O}(\varepsilon)$, which is the same order as $c_s^2$. In terms of $\sigma$, these functions are plotted in Fig. \ref{fig:Coefficients}. Then following the original step~\cite{Jaramillo:2020tuu}, we can define the energy inner product as
\begin{eqnarray}\label{EnergyScalarProduct}
    \langle u_1,u_2 \rangle_{\text{E}}=\int_0^1{\mathrm{d}\sigma}\Big[C_1\bar{\psi}_1\psi_2+C_2(\bar{\psi}_1\partial_{\sigma}\phi_2+\psi_2\partial_{\sigma}\bar{\phi}_1)+C_3\partial_{\sigma}\bar{\phi}_1\partial_{\sigma}\phi_2+C_4\bar{\phi}_1\phi_2\Big]\,,
\end{eqnarray}
where $u=(\phi\,,\psi)^{\text{T}}$ and $\psi=\partial_\tau\phi$ are introduced in Eq. (\ref{maineq}).
Finally, the energy norm is defined as
\begin{eqnarray}\label{energy_norm}
    \lVert u\rVert_{\text{E}}=\sqrt{\langle u, u\rangle_{\text{E}}}\, ,
\end{eqnarray}
which is strictly positive definite throughout the parameter regime considered in this work. The energy inner product \eqref{EnergyScalarProduct} and the energy norm \eqref{energy_norm} are related to the velocity $c_s$, which is a function of $\sigma$  [cf. Eq. (\ref{velocity})]. When $\varepsilon\to 0$, the phase velocity $c_s(\sigma)=1$, and the energy inner product along  with the energy norm revert to the original ones in~\cite{Jaramillo:2020tuu}.

\subsection{Pseudospectrum of EFT corrected black holes}
We aim to study the EFT corrected black holes' pseudospectrum, which is equivalent to $\sigma_\epsilon(L/i)$, since the QNM spectrum is the eigenvalue of $L/i$ [where $L$ is introduced in Eq. \eqref{maineq}]. For the computation of pseudospectrum, we introduce the Chebyshev-Lobatto grid combined with Clenshaw-Curtis quadrature to efficiently evaluate the integrals in Eq. \eqref{EnergyScalarProduct}. This approach involves computing a weighted sum of the function values at the grid points, leading to an inner product that incorporates a Gram matrix $\mathbf{G}$. This implies that \eqref{EnergyScalarProduct} is approximated by the following discrete version
\begin{eqnarray}
    \langle u_1,u_2 \rangle_{\text{E}}
    =\begin{pmatrix}
        \boldsymbol{\phi}_1^\star & \boldsymbol{\psi}_1^\star
    \end{pmatrix} \cdot \mathbf{G} \cdot
    \begin{pmatrix}
        \boldsymbol{\phi}_2 \\ \boldsymbol{\psi}_2
    \end{pmatrix}\, ,
\end{eqnarray}
where the asterisk denotes the Hermitian conjugate, and the Gram matrix $\mathbf{G}$ encodes the quadrature weights. The energy norm of an operator $A$ with matrix representation $\mathbf{A}$ is computed through
\begin{eqnarray}\label{matrixnorm}
    \lVert A\rVert_{\text{E}}=\lVert \mathbf{W} \cdot\mathbf{A}\cdot\mathbf{W}^{-1}\rVert_{2}\, ,
\end{eqnarray}
where $\mathbf{W}$ obtained from the Cholesky factorization $\mathbf{G}=\mathbf{W}^{\star}\cdot \mathbf{W}$ and $\lVert\cdot\rVert_2$ represents the matrix $2$-norm. This formulation allows us to evaluate the resolvent norm $\lVert R_{L/i}(\omega)\rVert_{\text{E}}$ at any complex frequency $\omega$  except at the QNM spectrum. The $\epsilon$-pseudospectrum $\sigma_\epsilon(L/i)$ then consists of all points $\omega$ satisfying $\lVert R_{L/i}(\omega)\rVert_{\text{E}} > \epsilon^{-1}$, as defined in Eq. \eqref{def of pseudo 1}. In practice, we generate the contour plots using
\begin{eqnarray}
    -\ln s^{\text{E}}_{\text{min}}(R_{\mathbf{L}/i}^{-1}(\omega))\equiv\ln(||R_{\mathbf{L}/i}(\omega)||_{\text{E}})\, ,
\end{eqnarray}
where $s^{\text{E}}_{\text{min}}(\mathbf{A})$ is the smallest of the “generalized singular values” for an arbitrary matrix $\mathbf{A}$, defined as
\begin{eqnarray}
    s^{\text{E}}_{\text{min}}(\mathbf{A}) \equiv s_{\text{min}}(\tilde{\mathbf{A}})\, ,
\end{eqnarray}
with $\tilde{\mathbf{A}}=\mathbf{W}\cdot \mathbf{A}\cdot \mathbf{W}^{-1}$ and $s_{\text{min}}$ is the smallest of the singular values. We note that the $2$-norm of a matrix is equal to its maximum singular value, and the $2$-norm of its inverse is equal to the reciprocal of its minimum singular value. Therefore $\lVert R_{\mathbf{L}/i}(\omega)\rVert_{\text{E}}> \epsilon^{-1}$ is equivalent to $-\ln s^{\text{E}}_{\text{min}}(R_{\mathbf{L}/i}^{-1}(\omega))>-\ln(\epsilon)$.  Further computational details can be found in~\cite{Jaramillo:2020tuu,Cao:2024oud,Chen:2024mon,Cao:2024sot}.

\begin{figure}[H]
    \centering
    \subfigure[]{\includegraphics[width=0.28\linewidth]{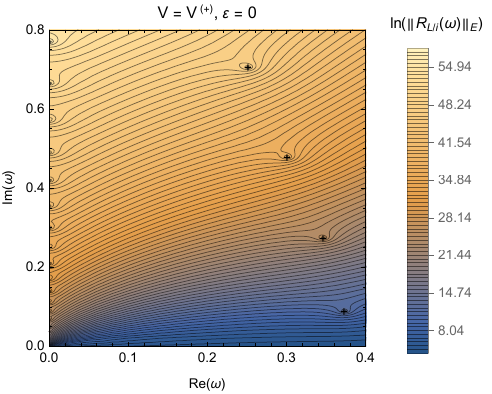}}  \hfill
    \subfigure[]{\includegraphics[width=0.28\linewidth]{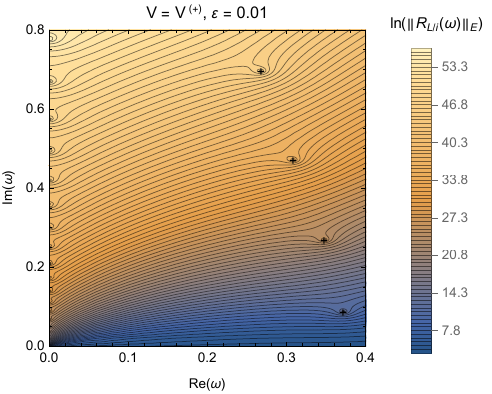}} \hfill
    \subfigure[]{\includegraphics[width=0.28\linewidth]{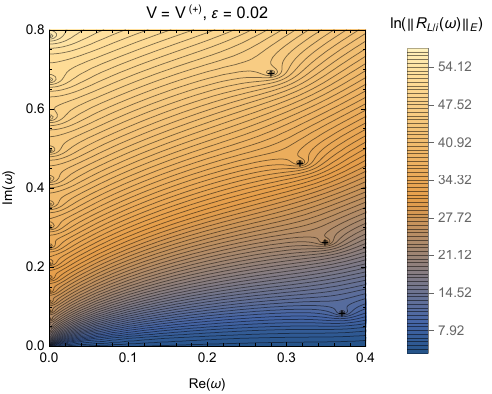}} \\
    \subfigure[]{\includegraphics[width=0.28\linewidth]{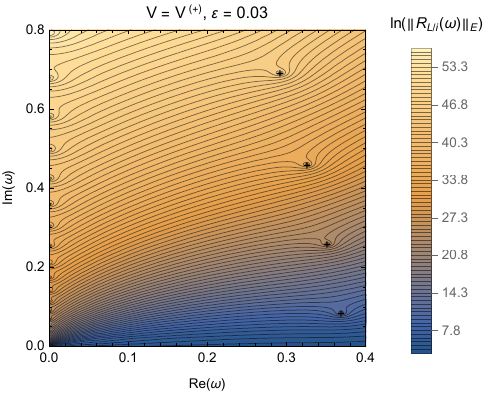}}  \hfill
    \subfigure[]{\includegraphics[width=0.28\linewidth]{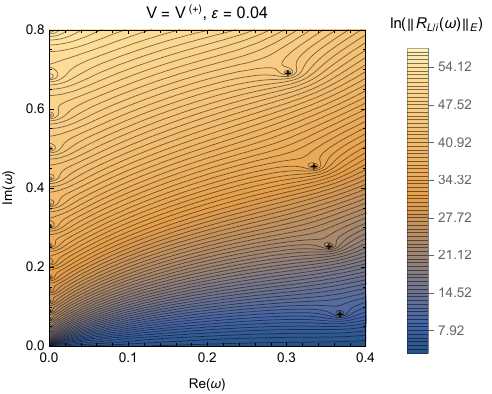}} \hfill
    \subfigure[]{\includegraphics[width=0.28\linewidth]{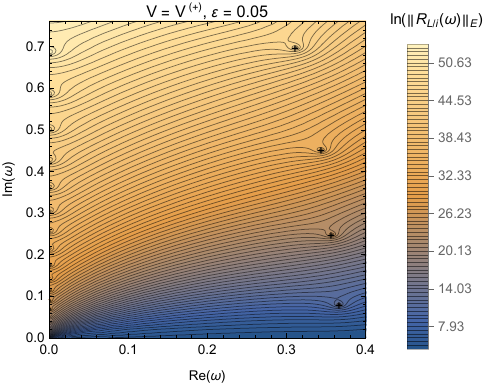}}
    
    \caption{(a-f): Pseudospectra for $\ell=2$ polar $(+)$ parity perturbations, with the parameter $\varepsilon$ varying from $0$ to $0.05$ in increments of $\Delta\varepsilon=0.01$. The calculation is performed on the Chebyshev-Lobatto grid with $N=100$. The QNM spectra are marked in $+$.}
    \label{fig:VpPseudo}
\end{figure}

\begin{figure}[H]
    \centering
    \subfigure[]{\includegraphics[width=0.28\linewidth]{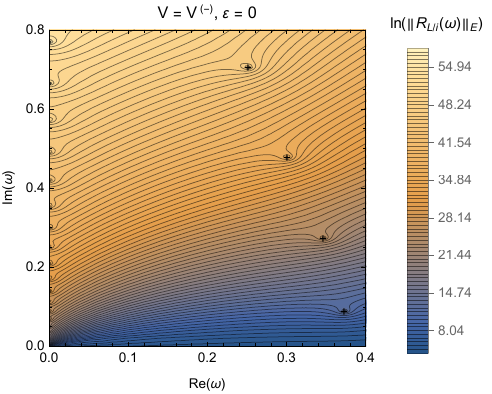}}  \hfill
    \subfigure[]{\includegraphics[width=0.28\linewidth]{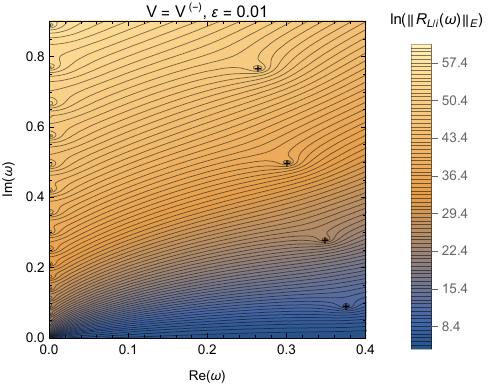}} \hfill
    \subfigure[]{\includegraphics[width=0.28\linewidth]{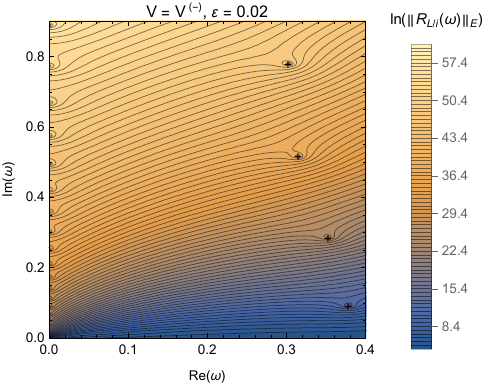}} \\
    \subfigure[]{\includegraphics[width=0.28\linewidth]{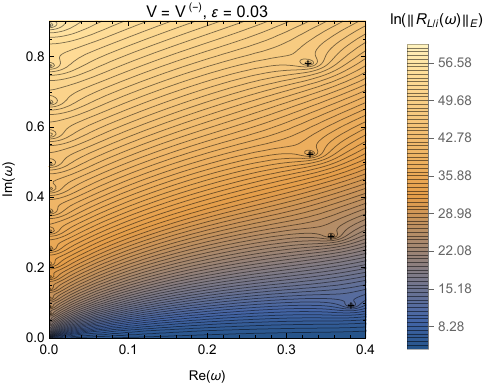}}  \hfill
    \subfigure[]{\includegraphics[width=0.28\linewidth]{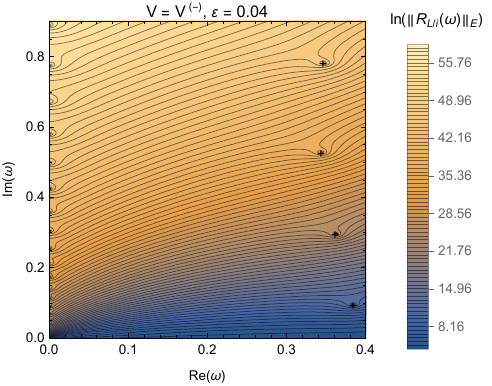}} \hfill
    \subfigure[]{\includegraphics[width=0.28\linewidth]{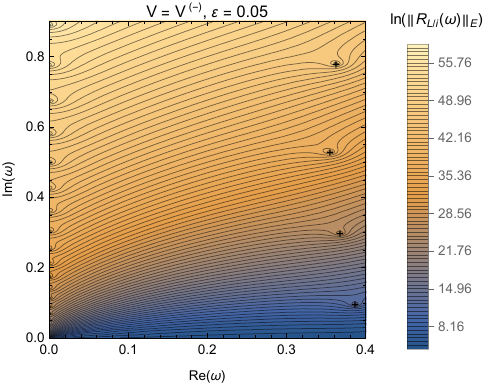}}
    \caption{(a-f): Pseudospectra for $\ell=2$ axial $(-)$ parity perturbations, with the parameter $\varepsilon$ varying from $0$ to $0.05$ in increments of $\Delta\varepsilon=0.01$. The calculation is performed on the Chebyshev-Lobatto grid with $N=100$. The QNM spectra are marked in $+$.}
    \label{fig:VmPseudo}
\end{figure}

Figures \ref{fig:VpPseudo} and \ref{fig:VmPseudo} are the pseudospectrum for polar $(+)$ and axial $(-)$ parity perturbations, respectively. These figures reveal two characteristics of the resolvent:
\begin{enumerate}
    \item When $\omega$ lies far from any QNM spectra, the energy norm of the resolvent $\lVert R_{\mathbf{L}/i}(\omega)\rVert_{\text{E}}$ grows monotonically with $\mathrm{Im}(\omega)$.
    \item When $\omega$ is in the vicinity of a QNM spectrum, $\lVert R_{\mathbf{L}/i}(\omega)\rVert_{\text{E}}$ tends to infinity as $\omega$ approaches such spectrum.
\end{enumerate}


For small values of $\epsilon$, the constant-$\epsilon$ contour forms a closed loop around the QNM spectrum. However, as $\epsilon$ increases slightly, the contour line expands significantly, developing an open structure that may extend to infinity. This indicates that perturbations of this size can lead to substantial shifts in the QNM spectra, which manifests that the QNM spectra are unstable. Each QNM spectrum has a critical threshold $\epsilon_c$ that distinguishes between closed and open contour lines in the pseudospectrum. For $\epsilon>\epsilon_c$, the spectrum becomes highly sensitive to perturbations. Thus, the stability of a QNM spectrum can be quantified by $\epsilon_c$, where a larger $\epsilon_c$ corresponds to greater robustness. 

\begin{figure}[H]
    \centering
    \subfigure[]{\includegraphics[width=0.48\linewidth]{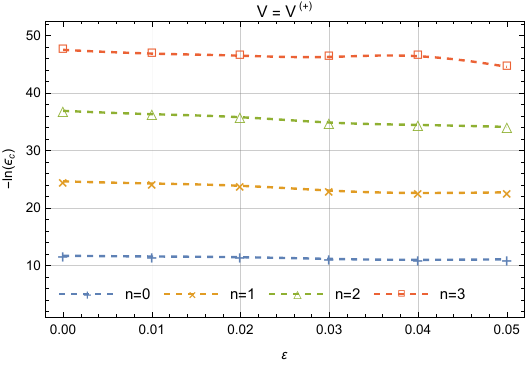}}  \hfill
    \subfigure[]{\includegraphics[width=0.48\linewidth]{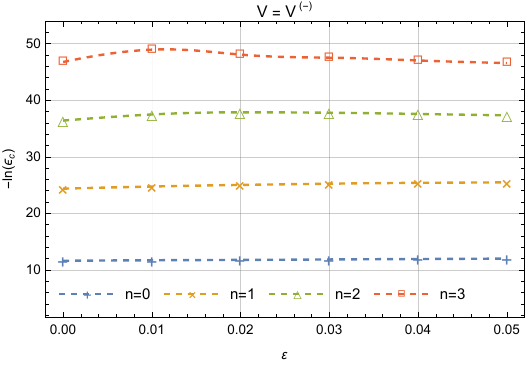}}
    \caption{$-\ln(\epsilon_c)$ for polar parity (a) pseudospectrum in Fig. \ref{fig:VpPseudo} and axial parity (b) pseudospectrum in Fig. \ref{fig:VmPseudo}. We vary the overtone number $n=0,1,2,3$ and the parameter $\varepsilon$ from $0$ to $0.05$ in increments of $\Delta\varepsilon=0.01$.}
    \label{fig:epsilon_c}
\end{figure}

We use the $-\ln s^{\text{E}}_{\text{min}}(R_{\mathbf{L}/i}^{-1}(\omega))$ data to calculate $-\ln(\epsilon_c)$ for the fundamental mode and the first to the third overtones, and we show our results in Fig. \ref{fig:epsilon_c}. We note that for the same $\varepsilon$, higher overtones have larger $-\ln(\epsilon_c)$, so they are less stable. For each mode, when $\varepsilon$ increases from $0$ to $0.05$, $-\ln(\epsilon_c)$  moves a little. For polar parity perturbations, $-\ln(\epsilon_c)$ decreases in most cases, except for the case $n=3$ where there is a local maximum at about $\varepsilon=0.04$. For axial parity perturbations, $-\ln(\epsilon_c)$ may be nonmonotonic, for $n=0,1$ it increases and for $n=3,4$ it first increases and then decreases. Although the instability of the black hole spectrum is not unique to GR and is a universal phenomenon, the relation between $\varepsilon$ and the magnitude of spectral instability is complex.

\section{Conclusions}\label{conclusions}
Under hyperboloidal framework, we can see from our results of the QNMs that the isospectrality is broken when dimension-$6$ terms are added in the action, which is consistent with previous works~\cite{Cardoso:2018ptl,Silva:2024ffz,deRham:2020ejn,Cano:2021myl}. We apply perturbation theory, a standard method in quantum mechanics, to calculate the $\mathcal{O}(\varepsilon)$ terms of the QNM spectra $\omega(\varepsilon)$ for fundamental mode and overtones. Based on Tables \ref{tab:Vp_kReIm} and \ref{tab:Vm_kReIm}, we find that higher overtones are more sensitive to the parameter $\varepsilon$ (or in other words, the EFT length scale $l$) in linear order. Furthermore, for higher overtones, the nonlinear effects are significant for smaller $\varepsilon$, which is the same as the results calculated by the phase-amplitude method~\cite{Silva:2024ffz}. Thus, for a fixed $\varepsilon$, there should be a threshold overtone number below which the linear approximation is effective, and the threshold overtone number decreases as $\varepsilon$ increases. Conversely, a threshold $\varepsilon$ exists for a fixed overtone number, decreasing as the overtone number increases. As for the time-domain waveforms, a small variation in $\varepsilon$ produces a small and continuous change in the time-domain waveform, rather than an abrupt one. Therefore, at least to the first order of $\varepsilon$, the time-domain waveform is stable. In order to provide a quantitative description, we also show the mismatch between the waveforms of $\varepsilon=0$ and distinct nonzero $\varepsilon$'s. Finally, the behavior of spectrum (in)stability characterized by the threshold $\epsilon_c$-pseudospectrum is not the same for different modes as $\varepsilon$ increases, i.e., it may increase, decrease or be nonmonotonic. We cannot conclude that the higher-dimensional curvature terms make the spectrum more or less stable, i.e., the effect is different for different modes.

The perturbation equation \eqref{tdeq} is established by assuming the variation of the Ricci tensor $\delta R_{\mu\nu}=0$ [cf. \cite{deRham:2020ejn}]. However, general perturbations do not have such restrictions and further studies are needed to investigate the $\delta R_{\mu\nu}\ne0$ case. In this case, the dimension-$4$ terms contribute to the perturbation equation~\cite{Held:2022abx} in the first order even in Ricci flat spacetime, and these terms make a difference in the QNM spectrum from that in GR~\cite{Antoniou:2024jku}. Moreover, the above calculation relies on first-order approximation on $\varepsilon$, and there should be further study on second-order approximation on $\varepsilon$ to investigate $\mathcal{O}(\varepsilon^2)$ behavior of QNMs. We note that the above analysis on the valid range of the first-order approximation on $\varepsilon$ is not accurate, as the resulting $\mathcal{O}(\varepsilon^2)$ terms of $\omega(\varepsilon)$ are not precise. In order to study when the first-order calculations are valid, one should use second-order approximation throughout the entire calculation process (field equation, background solution, perturbation equation and so on) and obtain the precise $\mathcal{O}(\varepsilon^2)$ terms of the QNM spectra. Besides, to confront gravitational-wave observations and place meaningful constraints on the parameter $\varepsilon$, it is essential to extend our spherical analysis to rotating black hole solutions and perform Bayesian parameter estimation with ringdown signals, following the methodology in~\cite{Silva:2022srr}. Finally, the QNMs and their (in)stabilities in a broader set of EFTs of gravity require further investigation.


\section*{ACKNOWLEDGMENT}
This work is supported in part by the National Key R\&D Program of China Grant No. 2022YFC2204603 and by the National Natural Science Foundation of China with Grants No. 12475063, No. 12075232 and No. 12247103. This work is also supported by the National Natural Science Foundation of China with Grants No. 12235019 and No. 11821505.
\appendix  

\section*{DATA AVAILABILITY}
The data that support the findings of this article are not publicly available upon publication because it is not technically feasible and/or the cost of preparing, depositing, and hosting the data would be prohibitive within the terms of this research project. The data are available from the authors upon reasonable request.

\section{The tortoise coordinate}\label{Appendix A}
\renewcommand{\theequation}{A\arabic{equation}} 
In this appendix, we provide more details about the tortoise coordinate. From Eq. (\ref{tortoise_coordinate}), the tortoise coordinate $r_{\star}$ satisfies
\begin{eqnarray}
    \frac{\mathrm{d}r_{\star}}{\mathrm{d}r}=\frac{1}{AB}= \Big(1-\frac{r_\text{h}}{r}\Big)^{-1}\Big[1+\varepsilon\Big(\frac{5}{8}\frac{M}{r}+\frac{5}{4}\frac{M^2}{r^2}+\frac{5}{2}\frac{M^3}{r^3}+\frac{5M^4}{r^4}+\frac{10M^5}{r^5}-\frac{88M^6}{r^6}\Big)+\mathcal{O}(\varepsilon^2)\Big]\, .
\end{eqnarray}
We can integrate this differential equation analytically, and the solution can be schematically written as~\cite{Silva:2024ffz},
\begin{eqnarray}\label{explicittor}
    r_{\star}(r)=r+r_\text{h}\ln(r-r_\text{h})+\varepsilon\delta r_{\star}(r)+\mathcal{O}(\varepsilon^2)\, ,
\end{eqnarray}
where
\begin{eqnarray}
    \delta r_{\star}(r)=r_{\text{h}}\Big[p_0+p_1 \ln(r_{\text{h}})+p_2\ln(r-r_{\text{h}})\Big]\, ,
\end{eqnarray}
with
\begin{eqnarray}
    p_0&=&\Big(\frac{5M^2}{2r_{\text{h}}^2}+\frac{5M^3}{r_{\text{h}}^3}+\frac{10M^4}{r_{\text{h}}^4}-\frac{88M^5}{r_{\text{h}}^5}\Big)\frac{M}{r}+\Big(\frac{5M^2}{2r_{\text{h}}^2}+\frac{5M^3}{r_{\text{h}}^3}-\frac{44M^4}{r_{\text{h}}^4}\Big)\frac{M^2}{r^2}\nonumber\\
    &&+\Big(\frac{10M^2}{3r_{\text{h}}^2}-\frac{88M^3}{3r_{\text{h}}^3}\Big)\frac{M^3}{r^3}-22\frac{M^2}{r_{\text{h}}^2}\frac{M^4}{r^4}\, ,\\
    p_1&=&-\frac{5M^2}{4r_{\text{h}}^2}-\frac{5M^3}{2r_{\text{h}}^3}-\frac{5M^4}{r_{\text{h}}^4}-\frac{10M^5}{r_{\text{h}}^5}+\frac{88M^6}{r_{\text{h}}^6}\, ,\\
    p_2&=&\frac{5M}{8r_{\text{h}}}+\frac{5M^2}{4r_{\text{h}}^2}+\frac{5M^3}{2r_{\text{h}}^3}+\frac{5M^4}{r_{\text{h}}^4}+\frac{10M^5}{r_{\text{h}}^5}-\frac{88M^6}{r_{\text{h}}^6}\, .
\end{eqnarray}
In order to study the singular behavior of $r_{\star}(\sigma)$ in the case $\sigma\rightarrow 0^{+}$ and $\sigma\rightarrow 1^{-}$, we study the singular behavior of $r_{\star}(r)$ in the case $r\rightarrow \infty$ and $r\rightarrow r_\text{h}$ respectively, i.e.,
\begin{eqnarray}
    r_{\star}(r)&\simeq& r+2M\ln r+\mathcal{O}(\varepsilon^2)\, ,\quad r\rightarrow\infty\, ,\label{rsim1}\\
    r_{\star}(r)&\simeq& r+r_{\text{h}}\ln(r-r_{\text{h}})+\varepsilon r_{\text{h}}p_2\ln(r-r_{\text{h}})+\mathcal{O}(\varepsilon^2)\, ,\quad r\rightarrow r_{\text{h}}\, .\label{rsim2}
\end{eqnarray}
In the limit $\varepsilon\rightarrow 0$, Eq. (\ref{explicittor}) reverts to the usual tortoise coordinate for the Schwarzschild black hole
\begin{eqnarray}
    r_{\star}(r)=r+2M\ln(r-2M)\, .
\end{eqnarray}

\section{The modifications of the effective potential}\label{Appendix B}
In this appendix, we give the modifications of the effective potential due to EFT correction, which can be written as
\begin{eqnarray}
    \delta V_{\ell}^{(+)} &=& \frac{1}{(r\Lambda_{\ell})^2} \sum_{i=1}^{10} v_{i\ell}^{(+)}(r) \Big(\frac{M}{r}\Big)^i\, , \label{deltaVa} \\
        \delta V_{\ell}^{(-)} &=& \frac{1}{r^2} \sum_{i=1}^{7} v_{i\ell}^{(-)} \Big(\frac{M}{r}\Big)^i\, .\label{deltaVb}
\end{eqnarray}

The polar parity potential coefficients are\footnote{Based on the effective potential in~\cite{deRham:2020ejn}, we obtain the coefficients of the modification terms, and the polar-parity coefficients are $1/4$ of that in~\cite{Silva:2024ffz}. Our expression for the effective potential yields a result identical to Fig. 1 in~\cite{Silva:2024ffz}}
\begin{eqnarray}
  v_{1\ell}^{(+)} &=& -\frac{5}{4}\lambda_{\ell}^2 (\lambda_{\ell} + 1)\, ,\nonumber \\
    v_{2\ell}^{(+)} &=& -\frac{5}{4}\lambda_{\ell}^2 (2\lambda_{\ell} + 5)\, ,\nonumber\\
    v_{3\ell}^{(+)} &=& -\frac{5}{4}\lambda_{\ell} (4\lambda_{\ell}^2 + 10\lambda_{\ell} + 9)\, ,\nonumber\\
    v_{4\ell}^{(+)} &=& -\frac{5}{4} (8\lambda_{\ell}^3 + 20\lambda_{\ell}^2 + 18\lambda_{\ell} + 9)\, ,\nonumber \\
    v_{5\ell}^{(+)} &=& \frac{5}{2} \Big[-8\lambda_{\ell}^3 - 20\lambda_{\ell}^2 - 18\lambda_{\ell} - 288\lambda_{\ell}^3 (\ell^2 + \ell - 6)/\Lambda_{\ell} - 9\Big]\, ,\nonumber \\
    v_{6\ell}^{(+)} &=& 176\lambda_{\ell}^3 + 116\lambda_{\ell}^2 - 90\lambda_{\ell} + 54\lambda_{\ell}^2[15 \ell^2(\ell+1)^2 - 336l(\ell+1) + 836]/\Lambda_{\ell} - 45\, ,\nonumber \\
    v_{7\ell}^{(+)} &=& 6 \Big(88\lambda_{\ell}^2 - 30\lambda_{\ell} + 15\lambda_{\ell} \Big[147 \ell^2(\ell+1)^2- 1304 \ell(\ell+1) + 2164\Big]/\Lambda_{\ell} - 15\Big)\, ,\nonumber \\
    v_{8\ell}^{(+)} &=& 36 \Big(-5 + 44\lambda_{\ell} + 3\lambda_{\ell} [1073 \ell(\ell+1) - 3988]/\Lambda_{\ell}\Big)\, ,\nonumber \\
    v_{9\ell}^{(+)} &=& 1584 + [194652 \ell(\ell+1) - 484560]/\Lambda_{\ell}\, ,\nonumber \\
    v_{10\ell}^{(+)} &=& 219888/\Lambda_{\ell}\, .  
\end{eqnarray}

The axial parity potential coefficients are
\begin{eqnarray}
    v_{1\ell}^{(-)} &=& -\frac{5}{8} \ell(\ell + 1)\, ,\nonumber  \\
    v_{2\ell}^{(-)} &=& -\frac{5}{4} (\ell^2 + \ell - 3)\, ,\nonumber  \\
    v_{3\ell}^{(-)} &=& -\frac{5}{2} (\ell^2 + \ell - 3)\, ,\nonumber  \\
    v_{4\ell}^{(-)} &=& -5 (\ell^2 + \ell - 3)\, ,\nonumber  \\
    v_{5\ell}^{(-)} &=& 1430 \ell(\ell + 1) - 8610\, ,\nonumber  \\
    v_{6\ell}^{(-)} &=& 41460 - 3332 \ell(\ell + 1)\, ,\nonumber  \\
    v_{7\ell}^{(-)} &=& -48192\, .
\end{eqnarray}

\bibliography{mainreference}
\bibliographystyle{apsrev4-1}

\end{document}